\documentclass{jpp}
\usepackage{graphicx}

\usepackage[utf8]{inputenc}
\usepackage[T1]{fontenc}
\usepackage{amsmath}
\usepackage{mathptmx}
\usepackage{url}
\usepackage{xcolor}
\usepackage{array}
\usepackage{multirow}

\newcolumntype{P}[1]{>{\raggedleft\arraybackslash}p{#1}}

\shorttitle{FLARE}
\shortauthor{H. Ji and others}

\title{The FLARE Facility}

\author{Hantao Ji\aff{1,2}
  \corresp{\email{hji@pppl.gov}},
  Jongsoo Yoo\aff{1}, Peiyun Shi\aff{1}, Euichan Jung\aff{2}, Kush Maheshwari\aff{2}, Adam Robbins\aff{2}, Sunghyun Son\aff{2}, Adam Stanier\aff{3}, Yang Ren\aff{1}, Sayak Bose\aff{1}, Dylan Corl\aff{1}, Keith Corrigan\aff{1}, Robert Cutler\aff{1}, William Daughton\aff{3}, Robert Ellis\aff{1}, Geoffrey Gettelfinger\aff{4}, Ronald Hatcher\aff{1}, Philip Heitzenroeder\aff{1}, Frank Hoffmann\aff{1}, Jonathan Jara-Almonte\aff{1}, Michael Kalish\aff{1}, Thomas Kozub\aff{1}, Enrique Merino\aff{1}, Weiguo Que\aff{1}, Benjamin Smith\aff{1}, John Wallace\aff{5}, Xin Zhao\aff{1},
  Sofia Avrutsky\aff{2}, Neal Crocker\aff{6}, Seth Dorfman\aff{6}, Yuka Doke\aff{7}, Corina Dunn\aff{2}, William Fox\aff{1}, Boting Li\aff{1}, Xiaocan Li\aff{3}, Clayton Myers\aff{1}, Shohgo Okazaki\aff{7}, Joshua Pawlak\aff{2}, Tongnyeol Rhee\aff{1}, Byonghoon Seo\aff{8}, Taiju Suzuki\aff{7}, Yang Zhang\aff{2},
  Stuart Bale\aff{9}, Amitava Bhattacharjee\aff{1,2}, Troy Carter\aff{10,6}, Chuanfei Dong\aff{11}, James Drake\aff{12}, Jan Egedal\aff{5}, Erik Gilson\aff{1}, Michiaki Inomoto\aff{7}, Jonathan Menard\aff{1}, Yasushi Ono\aff{7}, Stewart Prager\aff{1,2}, John Sarff\aff{5}, Hiroshi Tanabe\aff{7}, Masaaki Yamada\aff{1}
}

\affiliation{
\aff{1} Princeton Plasma Physics laboratory, Princeton, NJ 08540, USA
\aff{2} Department of Astrophysical Sciences, Princeton University, Princeton, NJ 08544, USA 
\aff{3} Los Alamos National Laboratory, Los Alamos, NM 07545, USA
\aff{4} Department of Physics, Princeton University, Princeton, NJ 08544, USA 
\aff{5} University of Wisconsin - Madison, Madison, WI 53706, USA
\aff{6} University of California, Los Angeles, Los Angeles, CA 90095, USA
\aff{7} University of Tokyo, Tokyo, Japan
\aff{8} Embry-Riddle Aeronautical University, Daytona Beach, FL 32114, USA
\aff{9} University of California, Berkeley, Berkeley, CA 94720, USA
\aff{10} Oak Ridge National Laboratory, Oak Ridge, TN 37830, USA
\aff{11} Boston University, Boston, MA 02215, USA
\aff{12} University of Maryland, College Park, College Park, MD 20742, USA
}

\begin{document}

\maketitle

\begin{abstract}
The Facility for Laboratory Reconnection Experiments (FLARE) has been constructed to study magnetic reconnection in multiple X-line regimes relevant to space, astrophysical, and fusion plasmas. Building upon the successful design of the Magnetic Reconnection Experiment (MRX), FLARE features a larger physical volume, stronger magnetic fields, and an independent ohmic heating drive to significantly extend the accessible parameter space, targeting Lundquist numbers up to $S \sim 10^5$ and normalized system sizes up to $\lambda \sim 10^3$. This paper details the facility's core engineering components, including the primary vacuum vessel, internal flux cores, highly segmented external coil systems, modular capacitor banks, and the safety interlock and control architecture. An initial diagnostic suite is presented, comprising high-resolution 2D magnetic probe arrays, triple Langmuir probes, a fully fiber-coupled interferometer, ion Doppler spectroscopy, and fast camera imaging. Initial operations demonstrate the device's experimental flexibility and reliability, successfully executing symmetric push-pull reconnection, spheromak merging, and asymmetric downstream configurations. Currently operating within \textit{Stage 2.5} with $S \sim 2,500$ and $\lambda \sim 60$ for anti-parallel reconnection, FLARE provides immediate access to the multiple X-line regimes. Planned hardware upgrades, advanced diagnostic additions, and integration with fully kinetic simulations will further expand its capabilities as it transitions into a collaborative user facility for the broader plasma science community.
\end{abstract}

\section{Introduction}

Magnetic reconnection~\citep{zweibel09,yamada10,ji22a} --- the rapid release of magnetic energy through topological rearrangement of magnetic field lines --- underlies many explosive plasma phenomena at all scales in the visible universe~\citep{ji11}. It plays a pivotal role in electron and ion heating, particle acceleration to high energies, plasma transport, and plasma morphology.  The impact on natural plasmas is enormous---from Earth's magnetospheric substorms to stellar flares to star formation to cosmic ray acceleration to the formation of extragalactic jets, as listed in Table \ref{t:astro} adapted from \cite{ji11}.  

The long history of magnetic reconnection research can be summarized in three major phases~\citep{ji22a}. The first phase began with magnetohydrodynamic (MHD) models, the simplest asymptotic theory, which is valid when the diffusion region thickness remains larger than the relevant ion kinetic scale. The historic MHD model~\citep{sweet58,parker57}, despite its confirmation in the laboratory at $S<10^3$~\citep{ji98}, fails in most applications where $S=10^8-10^{30}$~\citep{ji11} by orders of magnitude in predicting much slower reconnection rate of $R\approx S^{-1/2}$ than observation. The Petschek model~\citep{petschek64} does predict fast rates of $R\sim 0.01-0.1$ but requires a localized anomalous resistivity~\citep{biskamp01} which is difficult to observe~\citep{graham22,yoo24}.

The second major phase of research focused on physics beyond MHD as the origin of fast reconnection. This occurs when current sheet thickness is below ion kinetic scales~\citep{birn01} predicting the desired universal rate of $R\sim 0.1$~\citep{liu22}, including Hall effects~\citep[e.g.,][]{ma96} and anisotropic electron pressure~\citep[e.g.,][]{hesse99}. \textit{In-situ} space and laboratory measurements have confirmed these predictions on ion~\citep[e.g.,][]{mozer02,ren05} and electron scales~\citep[e.g.,][]{ji08,burch16,greess21}. A recent review~\citep{ji23} summarizes highlights from laboratory experiments on collisionless magnetic reconnection in direct relation with space measurements, especially by the Magnetospheric MultiScale (MMS) mission~\citep{burch16}.

However, these non-MHD mechanisms work only on kinetic scales, much smaller than the global scale of space and astrophysical plasmas~\citep{ji11}. A critical question arises: How can the MHD scale dynamics generate these thin layers, and when would this occur? The third major phase of the reconnection research began with the realization that the steady-state Sweet-Parker MHD model is unstable~\citep[e.g.,][]{biskamp86,shibata01,loureiro07} to the formation of secondary islands or plasmoids when $S>S_c \approx 10^4$~\citep{bhattacharjee09}, leading to multiple X-line reconnection. The thinner layers between plasmoids can further break up to form even thinner layers and so on, coupling MHD scales down to ion kinetic scales. The reconnection rate is independent of $S$~\citep[e.g.][]{bhattacharjee09,cassak09, uzdensky10} at $R\sim S_c^{-1/2} \approx 0.01$, fast enough to be relevant to observation. This effect applies to most space, solar, astrophysical and fusion plasmas. Plasmoids alter the reconnection dynamics fundamentally, including the possibility of accelerating particles to high energies~\citep{oka10,dahlin20,guo20,li2021,sironi25,yin26} as often observed. These new developments have been summarized in a recent \textit{roadmap} review~\citep{ji22a} in the context of long history of magnetic reconnection research with an emphasis on future prospects.

\begin{table}
\begin{center}
{\footnotesize
\begin{tabular}{p{15mm}p{37mm}p{45mm}p{30mm}}
Location & Plasma & Phenomena or Role & Multiple X-line Regime\\
\hline
Space & Earth's Magnetopause & Plasma Transport & Collisionless\\ 
& Earth's Magnetotail & Magnetic Substorm & Collisionless \\
\hline
Solar & Convection Zone & Solar Dynamo & Collisional\\
& Chromosphere & Plasma Heating and Jets & Hybrid/Collisional\\
& Corona & Solar Flare, Heating, Coronal Mass Ejections & Hybrid\\
& Solar Wind & Turbulent Dissipation & Collisionless\\
& Heliospheric Boundary & Dissipation, Transport & Collisionless\\
\hline
Milky Way & Protostellar Disk & Star/Planet Formation & Collisional \\
& X-ray binary Disk & Turbulent Accretion & Collisional \\
& X-ray binary Disk Corona & Disk Flare & Collisionless \\
& Crab nebula & $\gamma$-ray Flare & Hybrid \\
& Supernova &  $\gamma$-ray Burst & Collisional \\
& Magnetar & Magnetar Flare & Collisional \\
& Sgr A* & Sgr A* Flare & Collisionless \\
& Molecular Cloud & Star Formation & Collisional \\
& Interstellar Medium & Turbulent Dissipation & Hybrid \\
\hline
Extra-galactic & Active Galactic Nuclei Disk & Turbulent Accretion & Collisional \\
& AGN disk Corona & Disk Flare & Collisionless \\
& Radio Lobe & Plasma Heating & Hybrid \\
& Extragalactic Jet & Plasma Heating & Collisionless \\
& Galaxy Cluster & Plasma Heating/Cooling & Hybrid \\
\hline
Magnetic Fusion & Tokamak & Sawtooth Oscillation, Disruption, Edge Localized and Tearing Modes & Collisionless\\
& Reversed Field Pinch & Self-organization & Collisionless \\
& Spheromak & Self-organization & Collisionless \\
& Field Reversed Configuration & Formation & Collisionless \\
\hline
Inertial Fusion & Laser-driven & Externally Imposed or Self-generated & All  \\
& Z-pinch & Weakens Drive & Hybrid/Collisional \\
\end{tabular}
}
\end{center}
  \caption{Various space, solar, astrophysical, and fusion plasmas, in which magnetic reconnection is thought to play an important role, are listed with their approximate location shown in the multiple X-line regimes in the phase diagram~\citep{ji11}, also shown in Fig.~\ref{f:phase}.} \label{t:astro}
\end{table}

\begin{figure}
  \centering
  {\includegraphics[width=1\linewidth]{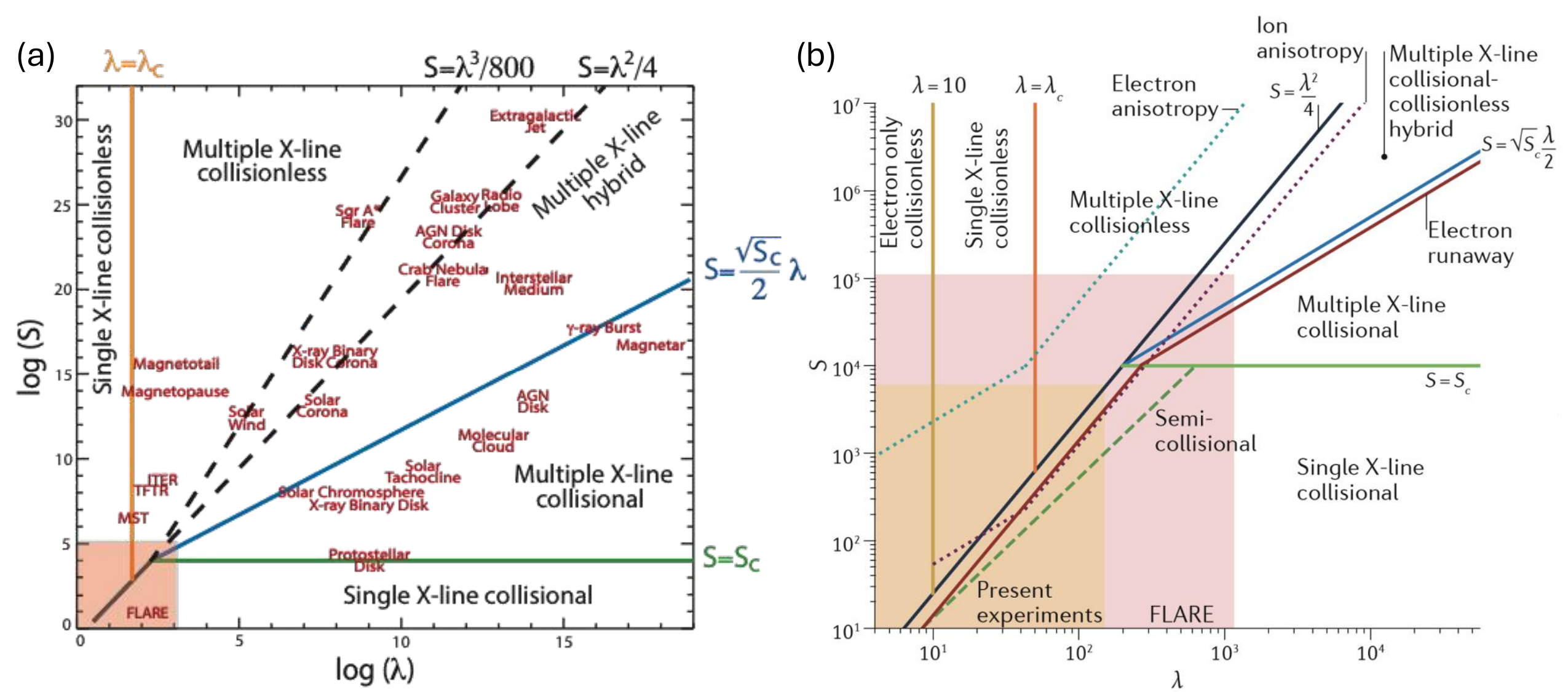}}
  \caption{A theoretical phase diagram for magnetic reconnection. If either the Lundquist number, $S$, or the normalized size, $\lambda$, is small, reconnection takes place in single X-line collisional regime or in collisionless regime without plasmoids. When both $S$ and $\lambda$ are sufficiently large, three multiple X-line regimes appear: \lq\lq collisional", \lq\lq collisionless", and \lq\lq hybrid" for collisional plasmoids with kinetic current sheets in between. (a) Reconnection phase diagram including heliophysical, astrophysical and fusion plasmas (see Table~\ref{t:astro}), adapted from \cite{ji22b} with the original form from \cite{ji11}. (b) Reconnection phase diagram including some recent refinements, adapted from \cite{ji22a}. The spaces covered by FLARE, as well as by previous experiments including MRX, are also shown.}
  \label{f:phase}
  \vspace{0mm}
\end{figure}

In order to provide laboratory access to the new reconnection phases or regimes, the FLARE (Facility for Laboratory Reconnection Experiments) device has been constructed and its design has been confirmed by a brief initial operation~\citep{ji18}. The new physics regimes~\citep{ji11} are characterized by dimensionless parameters that are orders of magnitude larger than currently available on devices such as the Magnetic Reconnection Experiment (MRX)~\citep{yamada97b} on which the FLARE design is based, and will therefore be more relevant to natural plasmas in the universe, as well as fusion plasmas. Specifically in comparison to MRX, $\lambda \equiv L/\rho_s$, the ratio of the system size, $L$, to the ion kinetic scale\footnote{This corresponds to the ion inertial length, $d_i$, for anti-parallel reconnection or the ion-sound gyroradius for guide field reconnection}, $\rho_s$, will increase from $\simeq 10^2$ to $\simeq 10^3$ and the Lundquist number\footnote{The factor of 4 in the $S$ definition arises from the assumption that the current sheet half-length is $L/4$ \citep{ji11}.}, $S \equiv \mu_0 L V_A/4 \eta _{Spitzer}$, the ratio of resistive diffusion time to Alfv\'en transit time ($\equiv L/V_A$; $V_A$ is the Alfv\'en speed), will increase from $\simeq 10^3$ to $\simeq 10^5$. This large advances in these two parameters enable access to critical new physics regimes as shown in the phase diagram (Fig.~\ref{f:phase}). 
It can yield a major advance by entering deeply into the important regime of multiple X-line collisionless plasmas and can enter the two new regimes: multiple X-line collisional and multiple X-line hybrid (a mixture of collisional and collisionless). Each of these regimes is important and relevant: for example, Table~\ref{t:astro} lists the approximate location within Fig.~\ref{f:phase} of the many venues for which reconnection is considered to be important to the observed explosive phenomena.  

The division of the $S-\lambda$ space into regions shown in Fig.~\ref{f:phase} is guided by current theory and simulations---which have been mostly limited to 2D studies.  FLARE will advance into these new regimes to discover whether the currently evolving theory is applicable and accurate, or whether a different division of physical regimes applies. In particular, some refinements to this basic phase diagram have been theoretically suggested in recent years, including the influence of electron pressure anisotropy~\citep{le15} and the potential existence of other sub-regimes which should be accessible by the FLARE device~\citep{huang13,cassak13,karimabadi13b,loureiro16,pucci17,bhat18}. Some of these refinements are included in Fig.~\ref{f:phase} (bottom panel) adapted from \cite{ji22a}. The goal of the FLARE project is to provide multiple communities a flexible yet robust, well-controlled and well-diagnosed laboratory platform to challenge these hypotheses and to explore for new discoveries.

Lastly, FLARE directly responds to the call by national and community reports: 2010 Plasma Decadal report~\citep{plasma07},
2016 DOE Frontier Plasma Science report~\citep{doe16}, and 
2020 DOE Community Planning Process report~\citep{plasma21}, as well as multiple community whitepapers~\citep[e.g.][]{ji23c} with more than 100 authors from more than 50 institutions around the world. In addition, FLARE is highly complementary to other ongoing reconnection experiments~\citep[e.g.,][]{ono96,olson16,lawrence09,moser12,nilson06,hare17,shi22,ji24} which have their own strengths.

The rest of this paper is organized in the following sections. In Sec.~2, the FLARE device is described including engineering design and construction of primary components. In Sec.~3, overview of initial diagnostics is given while in Sec.4 initial operational capabilities are illustrated with examples. The Sec.~5 provides descriptions of future upgrades and collaborative operations.

\section{Description of the Facility}

\subsection{Design Strategy and Parameters}

The goal of the FLARE project is to provide a laboratory access to magnetic reconnection at high Lundquist numbers and large normalized system sizes, enabling experimental studies in regimes directly relevant to space, astrophysical, and fusion plasmas. The FLARE design was based on that of the predecessor MRX device~\citep{yamada97b}. MRX has been at the center of the major advances in numerous areas of reconnection physics, for example, from collisional~\citep{ji98,kuritsyn07,jara-almonte16} to collisionless~\citep{ren05,yamada06,shi26}, from electron scale physics~\citep{ji08,ren08} to ion scale physics~\citep{yoo13}, from periodic to line-tied~\citep{oz11,myers15}, from zero guide field to finite guide field~\citep{tharp12,fox17,bose24}, from 2D to 3D physics~\citep{carter02a,ji04,dorfman13,yoo24}, from fully ionized to partially ionized plasmas~\citep{lawrence13}, and from symmetric to asymmetric~\citep{yoo14}. MRX has operated continuously, typically with more than 100 discharges on a single operation day, for 30 years without any major hardware failure, thereby demonstrating the reliability and flexibility of the design. 

\begin{figure}
  \centering
  \includegraphics[width=1.8in]{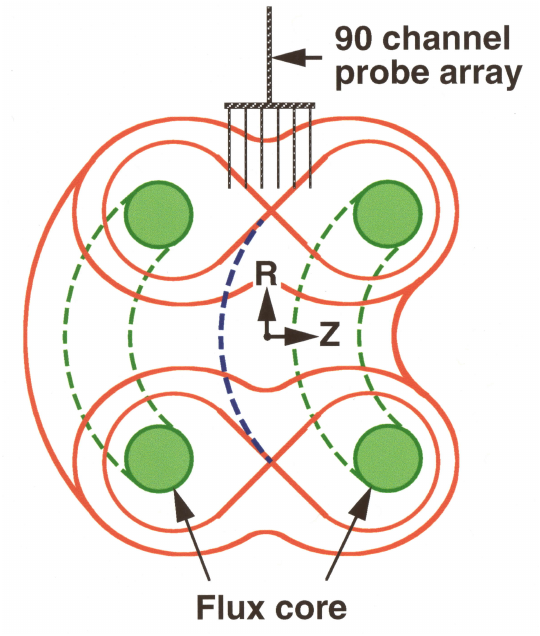}
  \raisebox{2mm}{\includegraphics[width=1.6in]{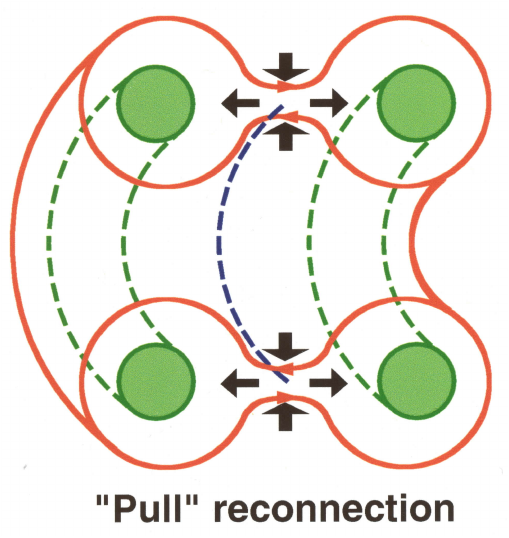}}
  \caption{(left panel) A quadrupole magnetic configuration formed by two flux cores. A typical 2D probe array is shown for illustration. (right panel) Magnetic reconnection is induced by reducing current in the flux cores by \lq\lq pulling" flux. Arrows indicate inflows and outflows of reconnection. Both panels are adapted from \citet{ji99}.
  \label{f:MRX}}
\end{figure}

Two key features made the MRX experiment unique and productive: 
(1) Control of the reconnection process by a pair of flux cores (Fig.~\ref{f:MRX}), accomplished by pulsing coil currents at different timings, and current levels under a variety of plasma conditions; and (2) Detailed diagnosis by a large array of pickup coils with sufficient spatiotemporal resolution to map out the 3D nature of reconnection. Figure~\ref{f:MRX} includes an illustration of the 90-channel probe array used in the early years of the MRX operations. 

FLARE builds on these successful features of MRX but provides a greatly extended parameter space and improved flexibility. The achievable $S$ and $\lambda$ with a finite guide field are given by
\begin{eqnarray}
S  & \equiv & \frac{\mu_0 L V_A}{4 \eta _{Spitzer}}= 1.1 \times 10^5 \left( \frac{L}{1.6 m} \right) \left( \frac{B_{rec}}{0.1T} \right ) \left( \frac{n}{10^{19} m^{-3}} \right)^{-1/2} \left( \frac{T_e}{30 eV}\right)^{3/2} \left(\frac{m_i}{m_H}\right)^{-1/2}\\
\lambda & \equiv & \frac{L}{\rho_s} = 1.0 \times 10^3 \left( \frac{L}{1.6 m}\right) \left( \frac{B_{guide}}{0.5 T}\right ) \left( \frac{T_e + T_i}{60 eV}\right)^{-1/2}\left(\frac{m_i}{m_H}\right)^{-1/2}.
\end{eqnarray}
where $B_{rec}$ and $B_{guide}$ are the reconnecting and guide field components, respectively. Here $L$ is system side corresponding to the surface-to-surface distance between flux cores. $m_i$ is ion mass and $m_H$ is hydrogen ion mass. For anti-parallel reconnection, a similar formula can be used for $\lambda$
\begin{equation}
\lambda \equiv \frac{L}{d_i} = 22 \left( \frac{L}{1.6 m}\right) \left( \frac{n}{10^{19} m^{-3}} \right)^{1/2} \left(\frac{m_i}{m_H}\right)^{-1/2}.
\end{equation}
A combination of a larger physical size, stronger magnetic field (both reconnecting and guide field components), and modestly higher temperatures can lead to the desired design values of $S$ and $\lambda$, simultaneously while allowing \textit{in-situ} measurements. 

\begin{table}
  \centering
  \vspace{-0mm}
  {\footnotesize
  \begin{tabular}{@{} ccc @{}}
    Parameters & MRX & FLARE \\ 
    \hline
    Device Diameter & 1.5 m & 3 m \\ 
    Device Length & 2 m & 3.6 m \\ 
    Flux Core Surface-to-surface Distance $L$ & 0.8 m & 1.6 m \\
    \hline
    Flux Core Major Radius $R_0$ & 0.375 m & 0.75 m \\ 
    Flux Core Minor Radius & 0.10 m & 0.15 m\\
\hline
    Guide Field at $R=R_0$ & 0.1 T & 0.5 T \\ 
    Ohmic Drive & No & 0.5 V-s \\ 
    Capacitor Bank Energy & $\sim30$ kJ & $\sim6.3$ MJ \\ 
\hline
Electron density ($10^{19}{\rm m}^{-3}$) & $0.1-5$  & $0.1-5$\\
Electron temperature (eV) & $3-15$ & $3-30$ \\
\hline
    $S$ (anti-parallel) & $600-1,400$ & $5,000-16,000$ \\ 
    $\lambda=L/d_i$ & $35-10$ & $100-30$ \\ 
\hline
    $S$ (guide field) & $2,900$ & $100,000$ \\ 
    $\lambda=L/\rho_s$ & 140 & 1,000 \\ 
  \end{tabular}}
  \caption{Key parameter comparisons between MRX and FLARE. \label{t:compare}}
\end{table}

Table~\ref{t:compare} summarizes key parameters of FLARE as compared with the existing MRX parameters. The device diameter is enlarged by a factor of 2 while the length is increased by a factor of 1.8. The length was determined by the requirements that it is long enough to accommodate the largest separation of two flux cores for $L=1.6$ m and there is enough space between vacuum vessel and flux cores for plasma breakdown and confinement, based on the ANSYS Maxwell calculations.  The guide field is increased by a factor of 5 in order to reduce the ion kinetic scale. An ohmic solenoid is added to drive and control reconnection more effectively and to provide more flexibility within the larger device. The ohmic drive can also help maintain and heat the plasma. A wide range of electron density is achievable by adjusting gas fill pressure. These advances, combined with better confinement due to the larger size and stronger fields, modest electron temperatures of 3-30 eV (for the desirable $S$) can be expected.

\begin{figure}
   \begin{center}
   \includegraphics[width=0.7\linewidth,angle=-90]{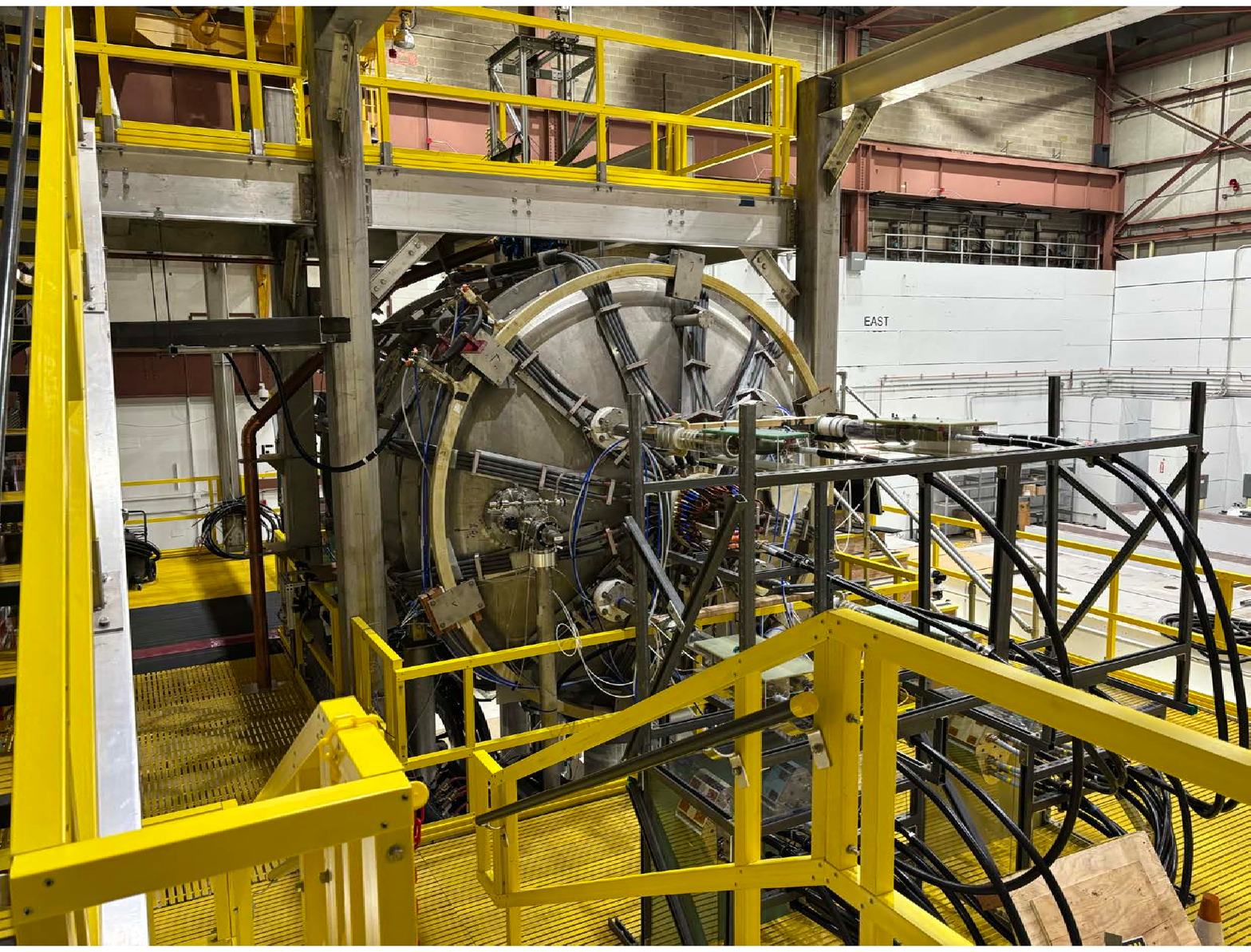}
   \caption{The FLARE device.}
   \label{fig:flare}
   \end{center}
\end{figure}

The FLARE facility consists of several major subsystems, including the vacuum vessel, internal flux cores, a center stack assembly, external coils, capacitor banks, data-acquisition system, control system, and machine-protection interlock system. These systems are described in turns below. A photo of the FLARE device is shown in Fig.~\ref{fig:flare}.

\subsection{Vacuum Vessel}
\begin{figure}
   \begin{center}
   \includegraphics[width=0.8\linewidth]{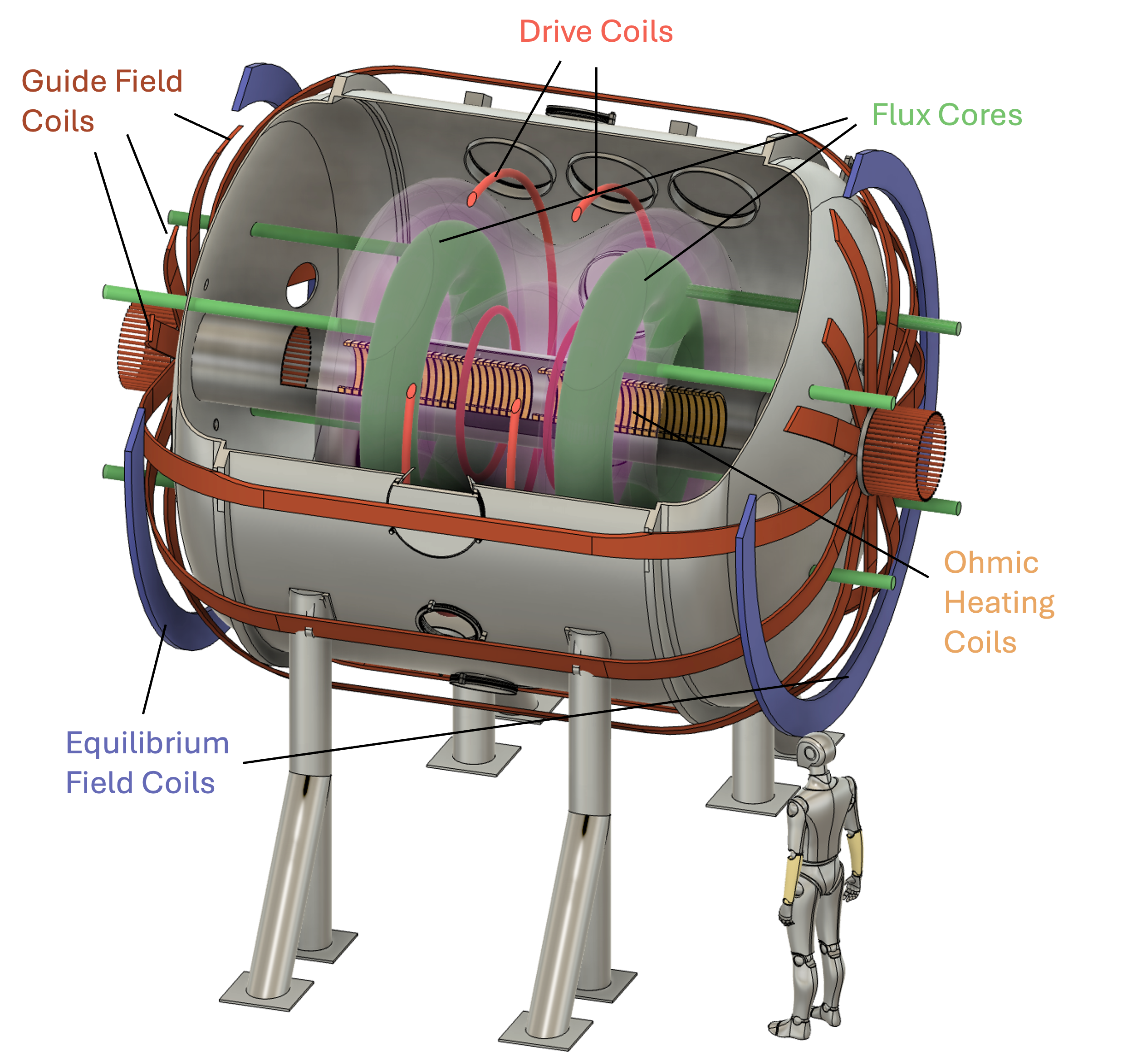}
   \caption{Three-dimensional rendering of FLARE with a cutaway view of the vacuum vessel and major coil systems. The primary confinement chamber is a fully welded Type 304 stainless-steel cylindrical vessel with a $2.46~\mathrm{m}$ outer length, $3.07$ m outer diameter, and $9.4~\mathrm{mm}$ wall thickness. Two endcaps complete the vessel, giving a total length of approximately $3.6$ m . Labeled components include the flux cores, ohmic-heating (OH) coils, guide-field (GF) coils, equilibrium-field (EF) coils, and drive coils (DC). Diagnostic and service ports distributed azimuthally around the vessel provide extensive access for plasma diagnostics and subsystem interfaces.}
   \label{fig:flare3d}
   \end{center}
\end{figure}

The FLARE vacuum vessel forms the primary confinement chamber for all plasma discharges. It is a fully welded structure fabricated from Type 304 stainless steel, chosen for its excellent vacuum performance, mechanical strength, and magnetic cleanliness (relative permeability $< 1.02$). All fabrication and welding conform to ASME B31.3 standards and Princeton Plasma Physics Laboratory (PPPL) quality-control procedures. Weld inspections satisfy the specified acceptance criteria, and all sealing surfaces are finished with a circumferential vacuum-grade machining pattern.

As shown in Fig.~\ref{fig:flare3d}, the vessel consists of a circular cylindrical shell with an outer length of approximately $2.46$ m (97 in.), outer diameter of $3.07$ m (120.75 in.), and wall thickness of $9.4$ mm (0.37 in.). With this thickness, the global $L/R$ magnetic diffusion time ($\tau = \mu_0 \sigma d r_0 / 2$; where $\mu_0$ is the vacuum permeability, $\sigma = 1.4 \times 10^6$ S/m is the conductivity, $d$ is the thickness, and $r_0 = 1.5$ m is the vessel radius) through the vessel wall is approximately 10 ms. Each end of the cylinder is sealed by a flat stainless-steel endcap weldment of the same wall thickness, providing a rigid pressure boundary and structural mount for internal and external components. Including the endcaps, the total vessel length is approximately $3.6$ m.

\begin{table}
\centering
\begin{tabular}{p{0.15\linewidth} p{0.15\linewidth} p{0.6\linewidth}}
\textbf{Location ($\theta^{\circ}$)} & \textbf{Flange Size} & \textbf{Primary Assignment} \\
\hline
$0^{\circ}$ & ISO 500-K & Main diagnostic port: magnetic probe arrays, Langmuir probes \\
$30^{\circ}$ & ISO 320-K & Fiber-optic interferometer \\
$60^{\circ}$ & ISO 500-K (3) & Future diagnostics \\
$90^{\circ}$ & ISO 500-K & Outer drive coil feedthrough \\
$120^{\circ}$ & ISO 320-K  & Ion Doppler spectroscopy (IDS) \\
$150^{\circ}$ & ISO 320-K & Fiber-optic interferometer \\
$180^{\circ}$ & ISO 320-F (2) \& ISO 500-K & Two turbomolecular pumps and a cryogetic pump \\
$210^{\circ}$ & ISO 500-K & Thomson scattering/LIF (injection) \\
$240^{\circ}$ & ISO 320-K & Fast camera \\
$270^{\circ}$ & ISO 500-K & Soft X-ray, Thomson scattering/LIF (collection) \\
$300^{\circ}$ & ISO 500-K (3) & Future diagnostics \\
$330^{\circ}$ & ISO 320-K & Thomson scattering/LIF (dump) \\
\end{tabular}
\caption{Summary of the FLARE vacuum vessel port allocations and flange sizes. Several azimuthal locations feature three ports to provide side access. (LIF=Laser-Induced Fluorescence.)}
\label{tab:port}
\end{table}

Diagnostic and service ports are distributed azimuthally around the cylindrical body at $30^{\circ}$ intervals. To accommodate the extensive diagnostic suite and operational hardware, the vessel utilizes a variety of port sizes and locations, as summarized in Table \ref{tab:port}. Most ports employ large stainless-steel vacuum flanges (nominal diameters of 21.65 in. and 14.57 in.) to house key diagnostics such as magnetic-probe arrays, Langmuir probes, and optical viewports [shown in Fig.~\ref{fig:ids}(a)] for the interferometer and ion Doppler spectroscopy (IDS), as well as Thomson scattering, Laser-Induced Fluorescence (LIF) and soft X-ray tomography in near future. Two 16.73-in. ports located on the bottom house the turbomolecular pumps. This symmetric port layout provides extensive mid-plane and end-plane access, enabling both two-dimensional and three-dimensional diagnostic coverage of the reconnection region. Additional ports on the endcaps support the center-stack assembly, two flux cores, vacuum gauges, and a piezoelectric valve for gas injection. 

\begin{figure}
    \centering
    \includegraphics[width=1\linewidth]{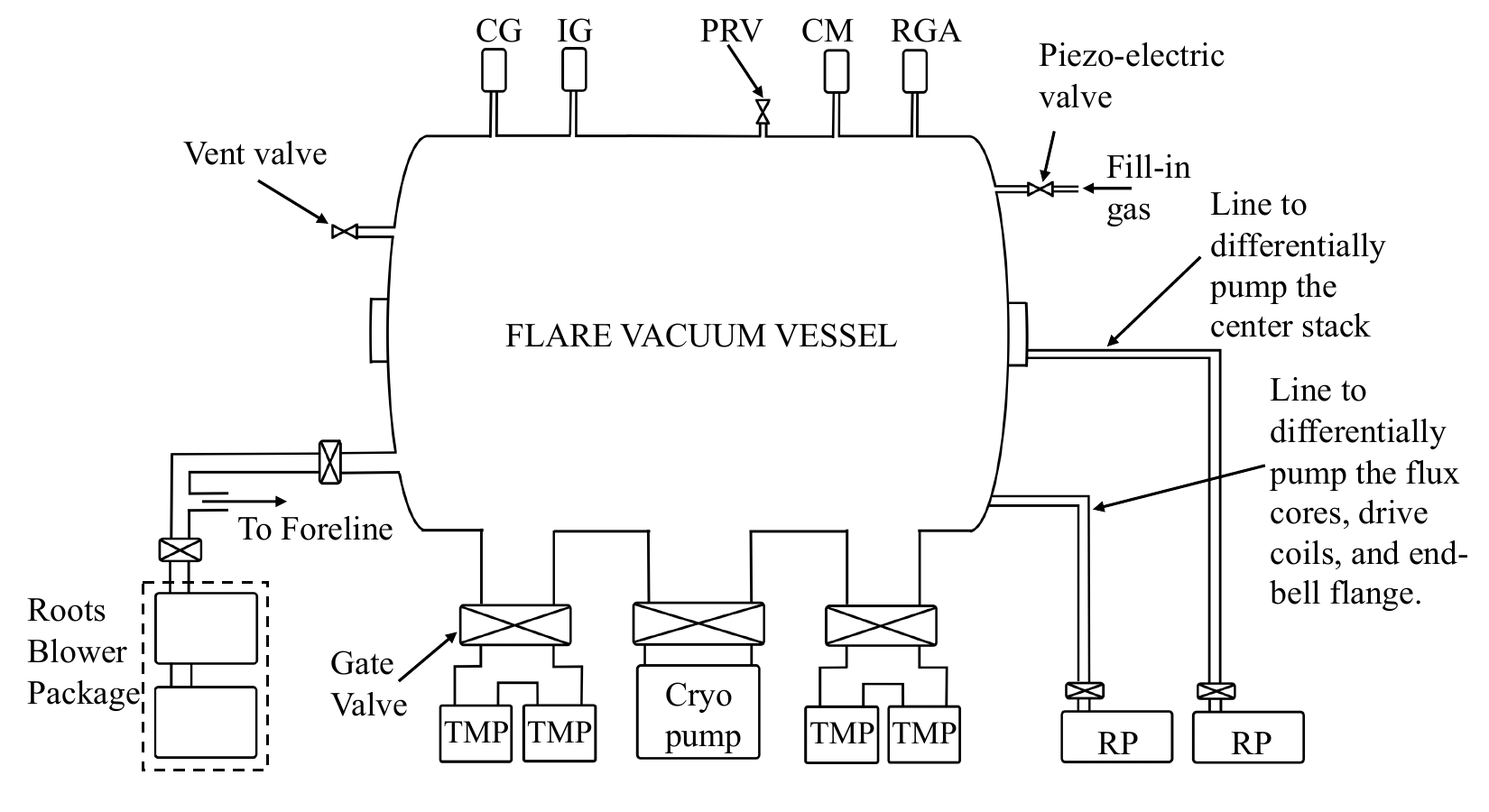}
    \caption{Diagram showing the main components of the FLARE vacuum system. The primary vacuum vessel is evacuated using a combination of four turbomolecular pumps (TMP) and a cryogenic pump, which are backed by a roots pump and a Rotary roughing pump. To manage gas loads from complex internal components, a couple of dedicated rotary pumps (RP) differentially pump the volumes within the flux cores, drive coils, the center stack, and end-bell flanges which has double viton o-rings. The schematic also illustrates the experimental gas injection line featuring a piezoelectric gas valve, alongside essential vacuum monitoring instrumentation including a Residual Gas Analyzer (RGA), an ion gauge (IG), convectron gauges (CG), and capacitance manometers (CM). FLARE is additionally equipped with a pressure release valve (PRV) and vent valve. The vacuum pumping system enables the facility to surpass its target base pressure of $1\times10^{-6}$ Torr, having already successfully achieved $5\times10^{-7}$ Torr during initial operations. See text for more details. }
    \label{fig:vacuum}
\end{figure}

To achieve the stringent high-vacuum conditions required for magnetic reconnection experiments, the FLARE facility employs a robust, multi-stage pumping architecture. The salient features of the FLARE vacuum system is shown  in Fig. \ref{fig:vacuum}. The primary evacuation of the approximately $1000\text{ ft}^3$ vacuum vessel is driven by a combination of up to four turbomolecular pumps (TMP) and a dedicated cryogenic pump. Each TMP has a pumping speed of 1500 L/s, while the cryogenic pump can evacuate at $10^{4}~\rm{L/s}$. These high-vacuum pumps are isolated from the main chamber by large pneumatic gate valves. Additionally there are  throttle valves between the TMPs and gate valves on turbo pumping lines which are not shown in the figure. The throttle valves  allow for precise conductance control during operations which enable us to control the effective pumping speed. The backing and initial roughing stages are handled by a Roots blower package, which consists of a Roots Blower pump and a rotary pump.

A critical engineering feature of the FLARE vacuum system is its extensive differential pumping manifold. Because the machine houses complex internal electromagnetic structures such as the flux cores, inner and outer drive coils, and the center stack assembly, these components contain internal volumes and dynamic seals that represent potential virtual leaks or outgassing sources. A differential pumping manifold actively evacuates these interstitial spaces, specifically targeting the flux core bodies, leg seals, terminations, and end bell flanges. This manifold ensures that localized outgassing or minor seal leaks bypass the primary chamber and do not compromise the ultra-high vacuum environment. In addition, the center stack assembly is differentially pumped using a separate rotary pump for greater flexibility. 

The system is fully instrumented to monitor pressure across all operating regimes, utilizing a network of convectron gauges (CG), capacitance manometers (CM), and a high-vacuum ion gauge (IG). Additionally, a Residual Gas Analyzer (RGA) is installed directly on the main vessel to monitor vacuum composition and assist in leak detection. Experimental working gas is introduced through a dedicated injection line regulated by a fast-acting piezoelectric gas valve. The target base pressure of FLARE is $1\times10^{-6}$ Torr, and a pressure of $5\times10^{-7}$ Torr has already been achieved using two turbomolecular pumps and a cryogenic pump mounted beneath the vessel, in combination with differential pumping of the two flux cores and the center stack. Further improvements in base pressure are planned to achieve higher electron temperature in future campaigns.

Two coordinate systems are used throughout this paper. For general machine descriptions, we employ a right-handed cylindrical coordinate system $(R,\theta,Z)$, where the positive $R$ direction points radially outward from the vessel axis, the positive $Z$ direction points toward the building west (toward the FLARE control room), and $\theta$ increases clockwise when viewed from the positive $Z$ direction. The angle $\theta = 0$ is defined along the vertical plane through the vessel axis pointing toward the top of the machine. We also use a local Cartesian coordinate system $(x,y,z)$, where $x = r\cos\theta$ and $y = r\sin\theta$, and $z$ is identical to that of the cylindrical system.

\subsection{Center stack assembly}
\begin{figure}
   \begin{center}
   \includegraphics[width=1\linewidth]{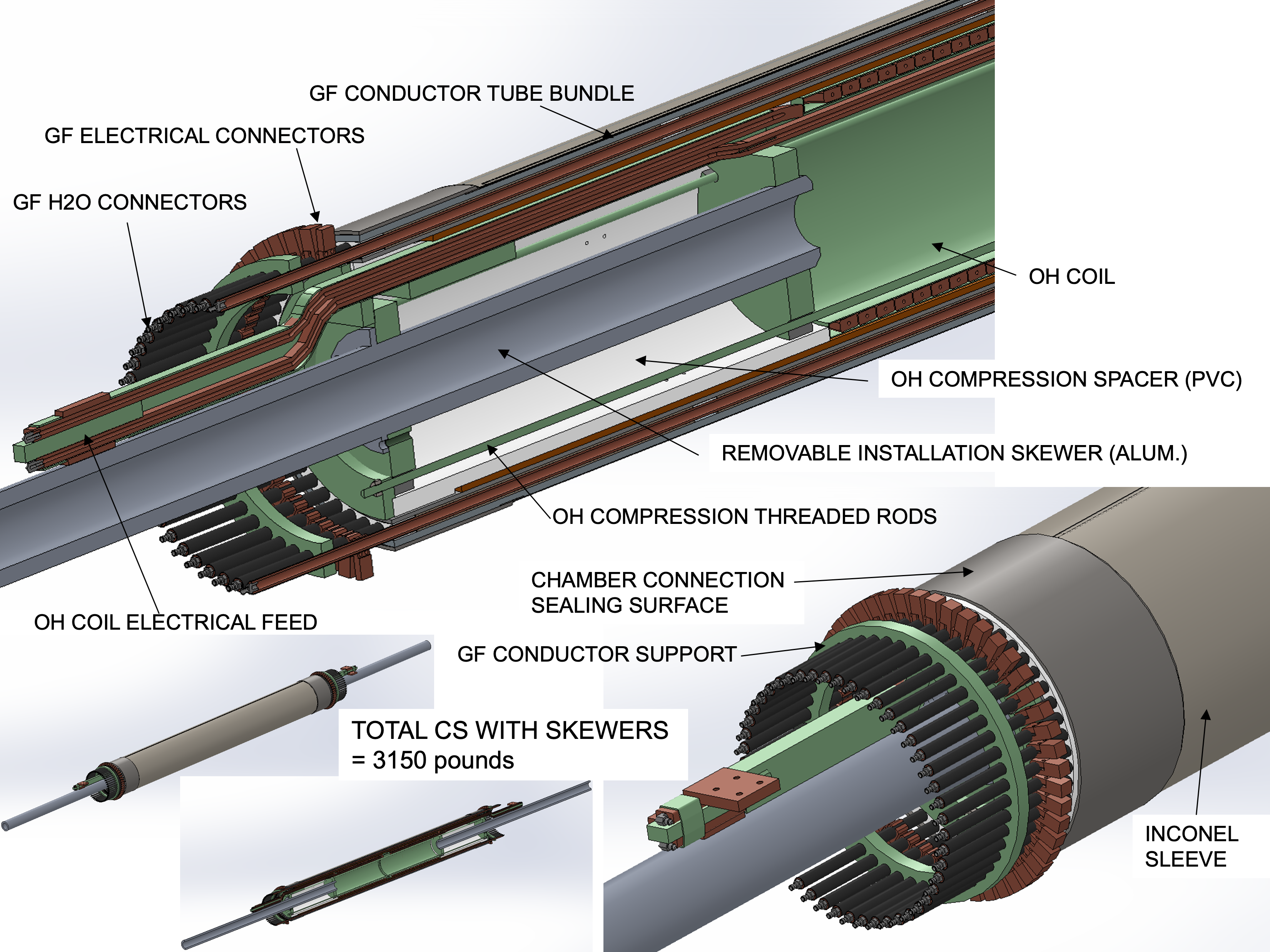}
   \caption{Three-dimensional view of the FLARE center-stack (CS) assembly. The CS is a 13-ft-long composite structure built from fiberglass and PVC tubes and reinforced by an outer Inconel sleeve. It houses two OH solenoid coils and 48 straight, water-cooled GF conductors. The OH coils are supported by PVC compression spacers and threaded rods, while the chamber interface includes sealing surfaces and structural support elements for vacuum operation. Removable 10-ft-long aluminum skewers are incorporated to facilitate installation and end-bell removal. The total assembly weight including the skewers is approximately 3150 lb.}
   \label{fig:cs}
   \end{center}
\end{figure}

The FLARE center stack (CS) assembly is a critical structural and electrical component that supports two coil systems while withstanding substantial electromagnetic and mechanical loads. The assembly is designed to withstand significant electromechanical stresses, including an applied torque of 36,000 ft-lbs and a total weight of approximately 3,150 lbs when including the installation skewers. To satisfy the requirement for a non-metallic housing, the design strategically utilizes a 13-foot-long composite structure fabricated from standard fiberglass and PVC tubes. The outer tube is made of a fiberglass tube with an 18.9 in. outer diameter and an 18 in. inner diameter. The inner tube is made of a PVC tube with a 16 in. outer diameter and a 15 in. inner diameter. This approach significantly reduced cost and fabrication lead time compared with specialty phenolic pultrusions. 

Space between inner and outer tubes houses the GF conductors and epoxy fill.  Fifteen gear like PVC spacer pieces hold the conductors in place.  These gears were aligned and then glued to the outer diameter of the inner PVC tube.  The conductor tubes were covered with shrink wrap, and then installed into the spacer gears.  This assembly was then slid inside the outer fiberglass tube.  Epoxy fill holes were drilled into the outer fiberglass tube to fill the spaces between spacer gears.  During the epoxy fill, the center stack was held vertical, and mixed epoxy was poured into the fill holes.  Heat tape and aluminum foil were wrapped around the outside to help with epoxy curing.  

The plasma-facing outer surface of the center stack is wrapped with a thin, 0.5 mm (0.020 in.) thick Inconel sleeve, which is seam welded in place, to provide mechanical robustness while maintaining a short magnetic-field penetration time.  To accommodate potential bowing of the underlying fiberglass (FRP) tubes, which have a typical outer diameter of 19.02 inches, the Inconel sleeve is designed with an inner diameter of 19.15 inches. Finally smooth surface stainless steel end-cuffs were welded to the Inconel.  These end-cuffs provide the O-ring vacuum sealing surface when assembled into the chamber. The length of this tube assembly is 4 meter.

As shown in Fig. \ref{fig:cs}, the center stack houses two Ohmic-Heating (OH) solenoid coils and a portion of the Guide-Field (GF) coil system. The OH coils are not interlocked with the GF coils allowing their possible replacements without disassembling the center stack. The GF system consists of 48 straight, water-cooled copper conductors, each with a cross-sectional area of 158~mm$^2$, designed to carry currents of up to 40 kA per conductor. These conductors require robust electrical isolation to hold off 20 kV for the overall in-out boundary and 416 V turn-to-turn. To achieve this, the conductors are sleeved in PVC heat-shrinkable tubing; thicker (0.040 in.) PVC is utilized for the in-out isolation, while thinner (0.004 in.) PVC is used for the individual turn-to-turn isolation. The entire conductor array is then fully encapsulated in ScotchWeld DP190 Gray epoxy adhesive. The OH solenoid is supported by PVC compression spacers and secured using dedicated threaded rods, ensuring precise alignment and mechanical stability under electromagnetic loading.

The CS assembly is reinforced using a multilayer material strategy, as illustrated in Fig.~\ref{fig:cs}. An external Inconel sleeve forms the outer boundary of the assembly, while internal structural integrity is achieved by bonding the inner and outer fiberglass tubes using Loctite Hysol E-120HP epoxy. Critical load-bearing interfaces utilize G-10 phenolic plates, selected for their excellent mechanical properties. Vacuum sealing to the chamber is provided by aluminum seal flanges, incorporating Delrin push rings and specialized O-ring grooves to ensure hermetic sealing. The ends of the assembly require a meticulous fabrication sequence to achieve vacuum leak-tight, full-penetration welds. Longitudinal under straps (approximately 1/16-inch thick) are first welded to each other. Next, 19.5-inch outer diameter cuffs (undercut to match the under strap) are welded to the under straps, and finally, the under straps are welded directly to the Inconel sleeve. The outer diameter of the internal center-stack assembly, including the Inconel sleeve, is 19.17 in. (48.7 cm). To protect the thin outer Inconel sleeve, the center stack provides enough strength to not sag while supporting the whole assembly. 

To facilitate installation of the 13-ft-long center stack into the vacuum vessel and allow removal of the end bells, the design incorporates removable aluminum skewers. These 10-ft-long skewers are fabricated from 5-in. Schedule-80 6061-T6 aluminum pipe and weigh approximately 167.6 lb each. Sockets for these skewers are also housed within the inner PVC tube. Structural simulations indicate a factor of safety (FOS) of 2.63, ensuring safe handling and mechanical reliability during installation and maintenance operations.

\subsection{Coil Systems}

The FLARE coil systems are also shown in Fig.~\ref{fig:flare3d} with parameters listed in Table.~\ref{t:summary}.  The vacuum vessel serves as the backbone of the device with all of the magnet systems and diagnostic equipment supported by its structure. Horizontally through the center of the vacuum vessel is a center bundle assembly that contains the ohmic heating (OH) coils and the center conductors for the guide field (GF) coils.  The removable OH coils are one-layer coils with a glass epoxy insulation system and extruded water cooled conductor.  Each of two OH coils is inserted from opposite ends of the center bundle providing a total of 50 turns.  The 48 center conductors for the GF coils consists of hollow water cooled copper conductor which is potted into the annulus of concentric fiberglass tubes in the center bundle assembly.  The GF coils are completed by connecting cables in groups of 4 to the GF conductor leads on the ends of the center bundle assembly and routing them on the outside of the vessel domes resulting in 12 segments with 48 turns in total. These cables are sized so that water cooling is not required. The equilibrium field (EF) coils are fabricated from extruded conductor (water cooled) with epoxy glass insulation.  Each coil is made from a double pancake assembly (two segments) which is mounted on either end of the vacuum vessel on the end dome flanges. Two pairs of drive coils (DC), with one pair (DC-I) located at the inner radius and the other pair (DC-O) at the outer radius, are used to drive and control reconnection, in addition to the drive by flux core PF coils and OH coils.

\begin{table}
  \centering
  {\footnotesize
  \begin{tabular}{@{} cccccccc@{}}
    Coil System & \# of & \# of Segments & \# of Turns & Radius & Voltage & Parallel & Peak Current \\ 
    & Coils & per Coil & per Segment & (m)  & (kV) & between & per Turn \\
    \hline
    Flux Core & 2 & 4 & 1& 0.75 average & 20 &  Segments & 135 kA \\ 
    Poloidal Field \\
    \hline
    Flux Core & 2 & 4 & 15 & 0.75 major & 20 & Segments & 62.5 kA \\ 
    Toroidal Field	&	&	&	&	0.13 minor \\
    \hline
    Guide Field & 1 & 12 & 4 & 0.22 inner & 20 & Turns & 40 kA \\
    & & & & 1.61 outer \\
    \hline
    Equilibrium Field & 2 & 2 & 8 & 1.50 & 1.4 & Coils & 13 kA \\ 
    \hline
    Ohmic Heating & 2 & 1 & 25 & 0.17 & 20 & Coils & 90 kA \\ 
    \hline
    Inner Drive Coil & 2 & 2 & 1 & 0.46 & 9 & Coils & 40 kA \\ 
    \hline
    Outer Drive Coil & 2 & 2 & 1 & 1.20 & 9 & Segments & 18.75 kA \\ 
  \end{tabular}}
  \caption{Summary of specifications of FLARE coil systems. \label{t:summary}}
\end{table}

\subsubsection{Flux Cores}
\begin{figure}
   \begin{center}
   \includegraphics[width=1\linewidth]{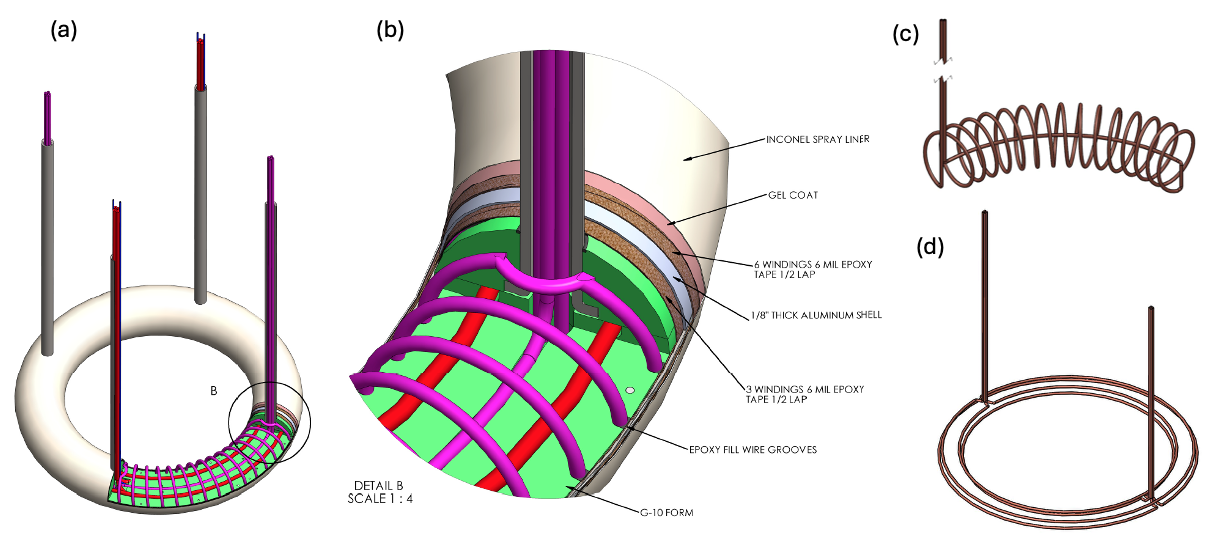}
   \caption{Flux-core geometry and winding structure in FLARE. (a) One flux core with toroidal geometry of major radius $75~\mathrm{cm}$ and minor radius $\sim 15.5~\mathrm{cm}$, supported by four structural legs. (b) Detail of the multilayer protection and insulation scheme surrounding the windings, including the G-10 fiberglass form, epoxy tape insulation, $0.125$ in. Al~3003 shell, and Inconel spray coating. (c) TF winding arrangement, using 60 turns divided into four 15-turn segments to reduce current requirements and toroidal-field ripple. (d) PF winding arrangement, consisting of four turns with radii of $65.4$, $72.5$, $77.5$, and $84.6~\mathrm{cm}$.}
   \label{fig:fc}
   \end{center}
\end{figure}

FLARE employs a pair of flux cores that generate the reconnecting magnetic fields, drive reconnection, and create plasmas. Located within the vacuum vessel, these two flux cores form the initial plasmas and are the key systems ensuring high accuracy in establishing the reconnection position. Each flux core contains two independent coil systems-the poloidal-field (PF) coils and the toroidal-field (TF) coils-and is designed to withstand large pulsed currents, with the peak currents for the TF and PF coils reaching 250 kA and 540 kA, respectively.

As shown in Fig.~\ref{fig:fc}(a), each flux core has a toroidal geometry with a major radius of $75~\mathrm{cm}$ and a minor radius of approximately $15.5~\mathrm{cm}$. Four structural \textit{legs} provide mechanical support for mounting the cores to the vacuum vessel. The coil windings, illustrated in Fig.~\ref{fig:fc}(b), are enclosed within G-10 fiberglass shells and protected by multiple layers of epoxy tape, a $0.125$"-thick aluminum (Al 3003) shell (segmented in 4 pieces), and a final Inconel-625 spray coating with a thickness of 0.14-0.19 mm (0.0055-0.0075 in.). This multilayer structure provides both dielectric insulation and protection from plasma exposure, while maintaining a short magnetic-field penetration time of order $\sim 10~\mu\mathrm{s}$.

Efficient reconnection drive requires minimizing the current rise time with desired peak currents in both the TF and PF windings. This requirement determines the number of turns and the degree of parallel segmentation used to reduce the total coil inductance. The TF windings act as a transformer whose inductive electric fields break down the working gas to generate plasma. Based on the MRX experience on the peak internal toroidal field strength of 1 Tesla and tolerable toroidal ripples, which can adversely affect plasma formation and damage flux core surface, a 60-turn configuration is adopted to reduce the required current per turn to 62.5 kA. As shown in Fig.~\ref{fig:fc}(c), the TF winding is divided into four segments which can be connected in parallel to reduce total coil inductance, each containing 15 turns with a return cancellation segment in the center in order not to generate a net current in the toroidal direction. This determines the peak TF current 62.5 kA $\times$ 4 = 250 kA. Two of the flux-core legs provide feedthroughs for the four sets of TF coil leads. The designed shortest rise time achievable is about 80 $\mu$s while the actually achieved rise time is about 100 $\mu$s (25\% longer) shown in Fig.~\ref{fig:vpic}(a) due to stray inductance especially from the busbar assembly in the circuit.

For the PF winding, shown in Fig.~\ref{fig:fc}(d), each flux core contains four turns with radii of $65.4$, $72.5$, $77.5$, and $84.6~\mathrm{cm}$, respectively. Two of the flux-core legs house the corresponding PF coil leads. The shortest rise time achievable is about 110 $\mu$s which corresponds to the actually achieved rise time of about 165 $\mu$s (50\% longer) shown in Fig.~\ref{fig:vpic}(a) due to stray impedance especially from the busbar assembly in the circuit. The larger rise time increase compared to the TF coils is due to the relatively smaller inductance of the PF coils. The peak PF coil current is determined by the ANSYS Maxwell calculations to achieve the desired reconnecting (poloidal) magnetic field of 0.1 Tesla at $R=0$ while taking into account of eddy currents in the vacuum vessel and the required current in the equilibrium field (EF) coil (see below) to locate the X-line at $R=0.75$ m. For the case of the surface-to-surface distance between flux cores $L=1.6$ m, the required total peak PF and EF coil currents are 540 kA and 176 kA, respectively. For $L=0.8$ m, the respective currents are 221 kA and 208 kA. However, unlike in MRX, we found that the global currents flowing in the FLARE plasma can have sizable effects in the TF and PF circuits, and thus the above calculations can only serve as a guidance rather than accurate predictions.

\begin{figure}
   \begin{center}
   \includegraphics[width=0.5\linewidth]{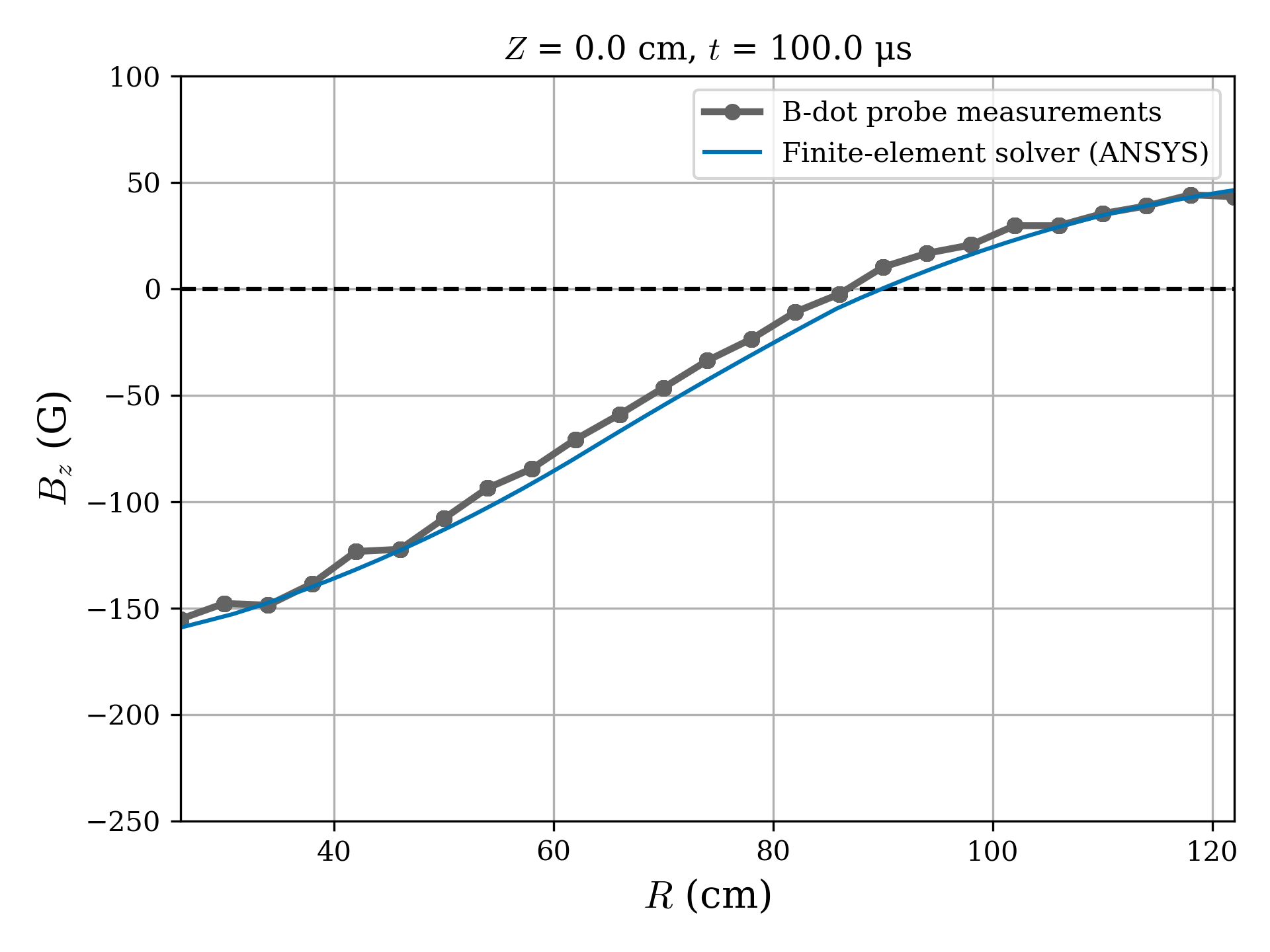}
   \caption{Comparison between the measured vacuum magnetic-field profile and a finite-element calculation for validation of the flux-core design and magnetic-field diagnostics. The figure shows a radial cut of the poloidal field $B_z$ at $Z=0$ and $t=100~\mu\mathrm{s}$ during a representative vacuum discharge. Gray symbols denote B-dot probe measurements, and the blue curve shows the corresponding ANSYS Maxwell result. The good agreement over the full radial range confirms that the flux-core system produces the intended magnetic-field structure and that the influence of vacuum-vessel eddy currents is accurately captured.}
   \label{fig:field_sim_comparison}
   \end{center}
\end{figure}

The TF and PF conductors are wound onto a three-plate G10 plastic form. The fabrication process of these three plates required special considerations. The 2.4 meter high feeder system requires the perpendicularity to be less than 1 mm, and three-dimensional spiral grooves require the surface outline dimension precision to be better than 0.5 mm. To avoid difficult machining helix and toroidal grooves directly on the G10 fiberglass with desired accuracy, a forming mold method is used. The molds for top and bottom plates have helix grooves for the TF windings and the middle plates have toroidal grooves for PF windings and TF return windings. Five-axis linkage CNC machining tools are used to machine these molds made of epoxy resin with high accuracies. Glass tapes are used inside the molds before G10 plates are made out of these molds.

The assembly and sealing of the flux cores require a meticulous multi-layer process to guarantee high-voltage insulation and vacuum integrity. During winding, fiberglass wool is packed into the cavities of the bottom and upper G-10 plates, in order to prevent plasma discharges under intense electric field at certain internal gas pressures. No thermal couples were placed within the flux core due to the same concerns. The plates are fixed with RTV108 silicone rubber once the cables are installed in the grooves. The four support legs, made of 316LN tubes with an outer diameter of 3 in. and a thickness of 0.5 in., are bolted into the bottom G-10 plate and sealed at the base using a special epoxy resin mixed with silicon powder. To prevent gas deflation into the vacuum vessel, all G-10 surfaces are wet-wrapped with fiberglass and a low-deflation-rate epoxy resin. Aluminum shells, lined with a polyimide insulation layer, are then installed over the surface. Crucially, these aluminum shells are divided into four separate pieces to prevent the formation of large induced eddy currents that would otherwise distort the magnetic field. The gaps between the aluminum shell segments are filled with epoxy putty to create a smooth surface, which is then wrapped in 8 half-lapped layers of self-vulcanized silicone rubber tape. Finally, multiple coats of Epotek 301-2 are applied until a 0.5 mm sealing layer is achieved, ensuring no gas emits from the flux core into the ultra-high vacuum environment. The volume within the flux core is held at rough vacuum to maintain experimental vacuum integrity and reduce pressure differential across the thin aluminum shell. High voltage and vacuum leakage tests are continuously carried out throughout the different stages of this fabrication process to ensure complete operational safety.

To cross-validate the flux-core design, the magnetic-field measurements, and the effect of vacuum-vessel eddy currents, we compared the measured vacuum field profile with a finite-element calculation. We used the ANSYS Maxwell adaptive-mesh finite-element transient solver to create a 2D cylindrical model of the FLARE vacuum vessel. For a representative vacuum shot, we input the measured poloidal field current and ANSYS computed the resultant eddy currents and magnetic field profile as a function of time. Figure~\ref{fig:field_sim_comparison} shows the resulting radial cut of the poloidal field $B_z$ at $Z=0$. As expected, the B-dot probe measurements agree well with the ANSYS Maxwell results over the full radial range. This agreement confirms that the flux-core system produces the intended magnetic-field structure, that the magnetic-field diagnostics are reliable, and that the influence of vacuum-vessel eddy currents is accurately captured.

\subsubsection{Drive Coils}
FLARE employs two sets of drive coils, the inner drive coils (DCI) and the outer drive coils (DCO), to provide an additional, adjustable reconnection electric field beyond that generated by PF and OH coils. Each coil set contains four total turns, arranged to produce a fast poloidal flux change that induces the required toroidal electric field in the plasma. These coils are designed to generate reconnection electric fields up to approximately $\sim 300~\mathrm{V/m}$.

The DCI coils consist of two identical two-turn coil assemblies supported by high-current feedthrough tubes mounted on endcap ports, enabling operation up to 80 kA. They are positioned close to the reconnection region, with a center radius of 46 cm and a 5 cm center-to-center spacing between the coil segments. The surface-to-surface separation between the two coil assemblies is approximately 50 cm, as shown in Fig.~\ref{fig:flare3d} (small red coils), which places them near the midplane for efficient flux coupling. To achieve inductance and rise-time characteristics comparable to those of DCO, the DCI coils use a half-parallel, half-series connection, in which the two turns on each side are connected in series, and the two resulting branches are connected in parallel.

The DCO coils form the larger-radius pair, with a center radius of 1.2 m, a 3.8 cm center-to-center spacing between segments, and a surface-to-surface spacing of approximately 60 cm. The DCO coils are supported by 12 structural mounts welded around the vessel, ensuring mechanical stability during high-current operation (up to 75 kA). The recommended configuration for DCO operation is an all-parallel connection of all four turns, which minimizes inductance and maximizes current. This arrangement allows the DCO system to impose a poloidal flux perturbation on a time scale comparable to that of the DCI coils.

Together, the DCI and DCO coils provide a flexible inductive-drive capability that allows FLARE to tailor both the magnitude and spatial distribution of the reconnection electric field without significantly perturbing the background magnetic-field geometry. This capability enables controlled studies of fast-reconnection onset, current-sheet thinning, and the parametric dependence of the reconnection rate~\citep{shi26,jara-almonte16}. The design of both drive-coil sets has been completed, and their fabrication and installation are planned for the near future, which will further expand the operational capabilities of FLARE.

\subsubsection{Guide Field Coil}

\begin{figure}
   \begin{center}
   \includegraphics[width=0.6\linewidth]{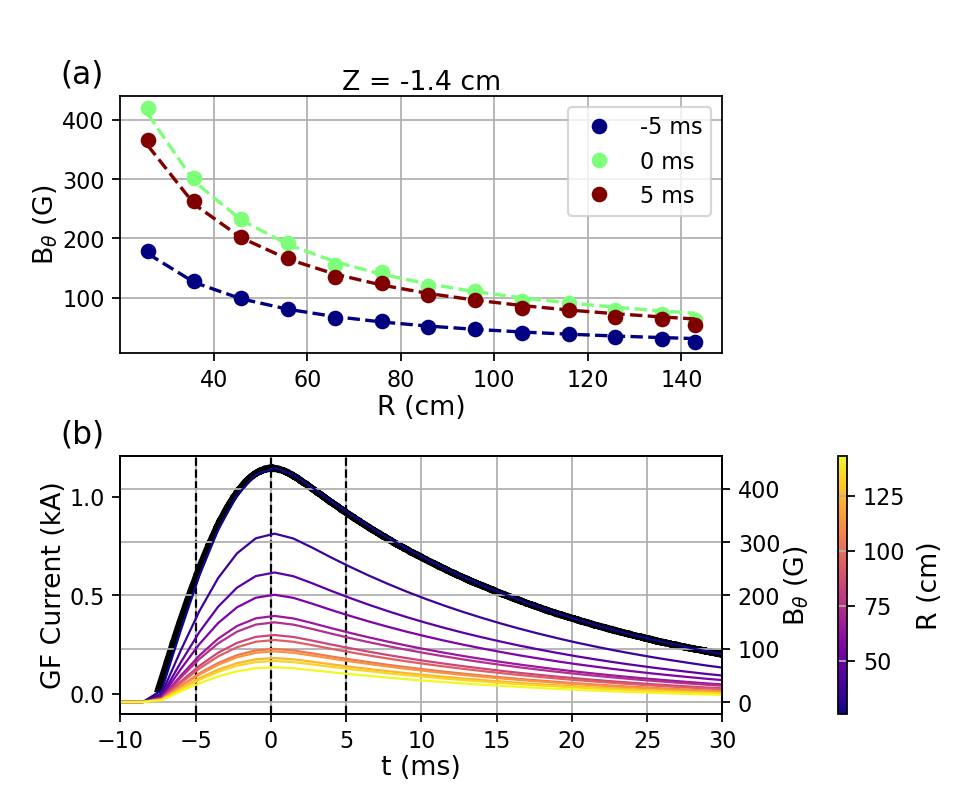}
   \caption{Characterization of the FLARE guide-field (GF) coil system using in-vessel magnetic-field measurements obtained with a Metrolab THM1176-LF magnetometer during vacuum discharges. (a) Radial profiles of the toroidal field $B_{\theta}$ measured at $Z=-1.4~\mathrm{cm}$ for three representative times, showing the expected $1/R$ dependence. (b) Temporal evolution of the GF coil current (black curve, left axis) and the measured $B_{\theta}$ waveforms (colored curves, right axis) at different radii. The close temporal tracking between the current and magnetic-field signals indicates negligible effects from vacuum-vessel eddy currents during GF operation.}
   \label{fig:gf}
   \end{center}
\end{figure}

The guide-field (GF) coil system provides a controllable toroidal magnetic field and, in terms of the local reconnection geometry, sets the out-of-plane or \textit{guide-field} component. As shown in Fig.~\ref{fig:flare3d}, the GF system consists of 48 total turns distributed across 12 equally spaced locations around the vacuum vessel. The current loops are closed through copper tubes with an outer diameter of 0.75 in. in the center-stack assembly, forming an azimuthally symmetric toroidal-field ($B_{\theta}$) coil set. These conductors carry 40 kA for 40 ms.  Since a 20 kV potential is used to drive the GF coil current, the turn-to-turn insulation is designed to hold off 500 V. Because the coil set fully encircles the vessel, the resulting magnetic field is predominantly toroidal and follows the expected radial dependence $B_{\theta}(R) \propto 1/R$.

To characterize the GF field and cross-check the effect of the vacuum vessel on field penetration, we measured the magnetic field inside the vessel using a compact three-axis Hall-effect magnetometer (Metrolab THM1176-LF) during dedicated vacuum discharges. As shown in Fig.~\ref{fig:gf}(a), the measured field profile exhibits the expected $1/R$ dependence over a wide radial range. Figure~\ref{fig:gf}(b) compares the temporal evolution of the GF coil current with the measured magnetic-field waveforms at different radii. No appreciable time delay is observed between the coil-current waveform and the measured field, indicating that the effect of vessel eddy currents on the GF field is minimal. This behavior is consistent with the relatively slow GF-bank discharge compared with the vessel magnetic diffusion time.

At present, the maximum applied guide field at the geometric center of the machine ($R = 0.75~\mathrm{m}$) is approximately $0.16~\mathrm{T}$. We plan to increase this value to $0.5~\mathrm{T}$ in the future. The GF system can operate with either positive or negative toroidal-field polarity, providing flexibility in establishing a wide range of guide-field conditions. Strong guide fields are important for achieving large $\lambda$, since increasing the guide field $B_g$ reduces the ion sound Larmor radius $\rho_s$ and enables access to new reconnection regimes.

\subsubsection{Equilibrium Field Coils}

\begin{figure}
   \begin{center}
   \includegraphics[width=0.75\linewidth]{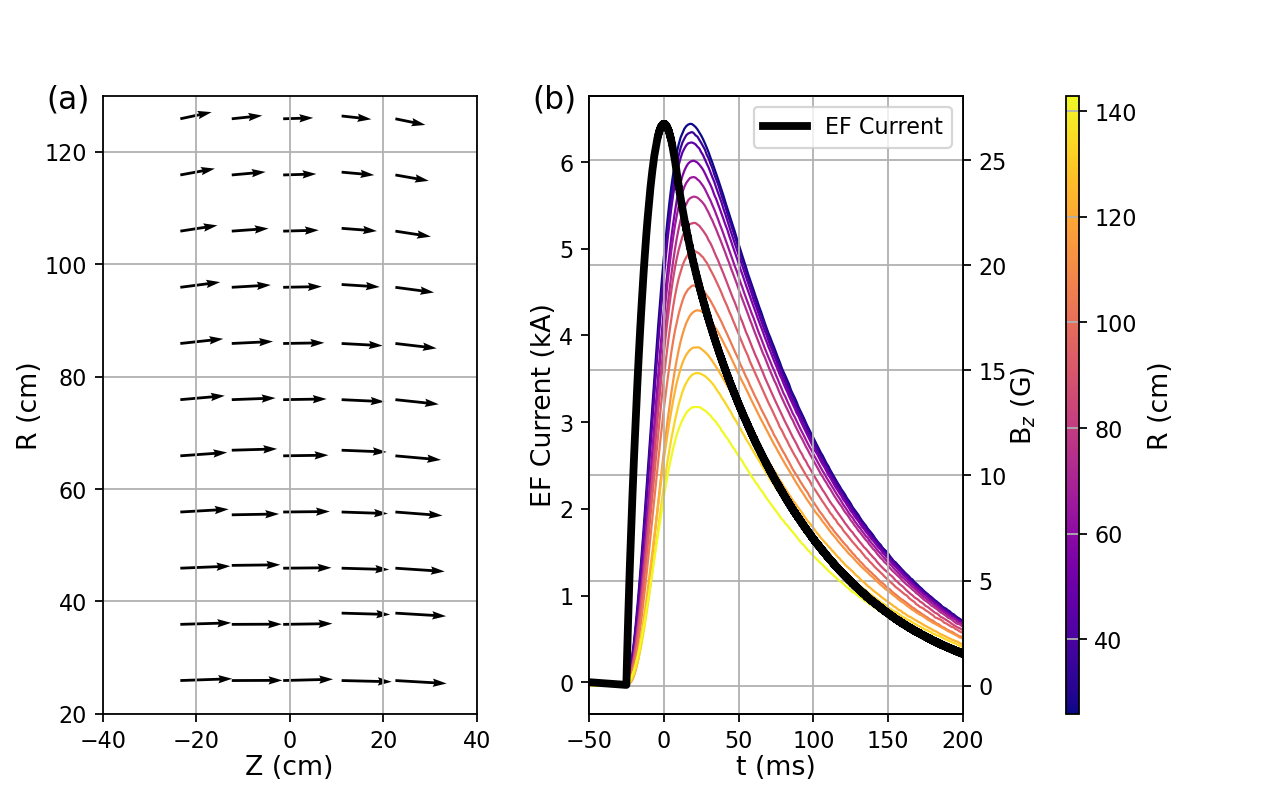}
   \caption{Characterization of the FLARE equilibrium-field (EF) coil system during vacuum discharges using a compact three-axis Hall-effect magnetometer (Metrolab THM1176-LF). (a) Two-dimensional measurement of the magnetic-field direction in the $R$–$Z$ plane, demonstrating the expected spatial structure of the EF field. (b) Temporal evolution of the EF coil current (black, left axis) and the measured axial field $B_Z$ (colored curves, right axis) at different radii. The substantial lag of the measured field relative to the coil current confirms significant magnetic diffusion through the conducting vessel wall.}
   \label{fig:ef}
   \end{center}
\end{figure}

The Equilibrium Field (EF) coil system consists of two coil sets on each side of FLARE, as shown in Fig.~\ref{fig:flare3d}. Each EF coil assembly has 8 turns with an average turn diameter of approximately $3~\mathrm{m}$, allowing the system to generate a relatively uniform axial magnetic field ($B_Z$) across FLARE, together with a smaller radial component ($B_R$). The primary function of the EF coils is to control the radial location of the current sheet by adjusting the location of $B_Z=0$ and providing the radial force balance required for equilibrium.

To characterize the EF field, we carried out two-dimensional magnetic-field measurements over a wide range of $R$ and $Z$ using the same compact three-axis Hall-effect magnetometer (Metrolab THM1176-LF) during dedicated vacuum discharges. As shown in Fig.~\ref{fig:ef}(a), the measured field pattern generally agrees with the expected spatial structure of the EF coil system, confirming that the coils produce the intended large-scale equilibrium field. However, the temporal response of the measured field differs substantially from that of the coil current. Figure~\ref{fig:ef}(b) shows that the magnetic field inside the vessel rises more slowly than the EF coil current and reaches its maximum significantly later, indicating strong effects from eddy currents in the conducting vacuum vessel.

Because the EF coils are mounted externally on the vacuum vessel and are designed to generate relatively slowly varying large-scale fields, eddy-current effects lead to both attenuation and temporal distortion of the applied field. In particular, the peak internal field occurs roughly $10~\mathrm{ms}$ after the peak coil current, and the measured field strength is somewhat smaller than the vacuum-field expectation. These effects must be taken into account when programming the EF waveform, especially in experiments requiring precise timing of the current-sheet position. Despite this delay, the EF system provides robust, shot-to-shot control of the current-sheet location. At present, the maximum field strength produced by the EF coils near the center of FLARE ($Z=0$, $R=0.75~\mathrm{m}$) is about $300~\mathrm{G}$.

\subsubsection{Ohmic Heating Coils}
The Ohmic Heating (OH) coil system consists of two solenoid coil assemblies, one on each side of FLARE, located inside the center-stack structure as shown in Fig.~\ref{fig:flare3d}. Each assembly contains a 1.01 m-long solenoidal winding with 25 turns in one layer, and the total length including the lead sections is approximately 2.71 m. The outer diameter of each turn is about 0.37 m. The OH coils are designed to handle 100 kA for 3 ms and provide a poloidal flux of 0.53 V-s with the maximum one-turn voltage of 800 V when they are connected in parallel and powered by the OH capacitor bank of 20 kV.

The primary purpose of the OH coils is to provide additional electron heating by driving extra toroidal plasma current through transformer action. In this sense, the OH coils supply an auxiliary reconnection drive, producing a magnetic field dominated by its $B_Z$ component, which has the opposite sign to the field generated by the EF coils. At their maximum current, the OH coils produce a field strength of approximately 100 gauss at $(R,Z) = (0.75~\text{m}, 0)$.

\subsection{Capacitor banks}

\begin{figure}
  \centering 
  \includegraphics[width=3in]{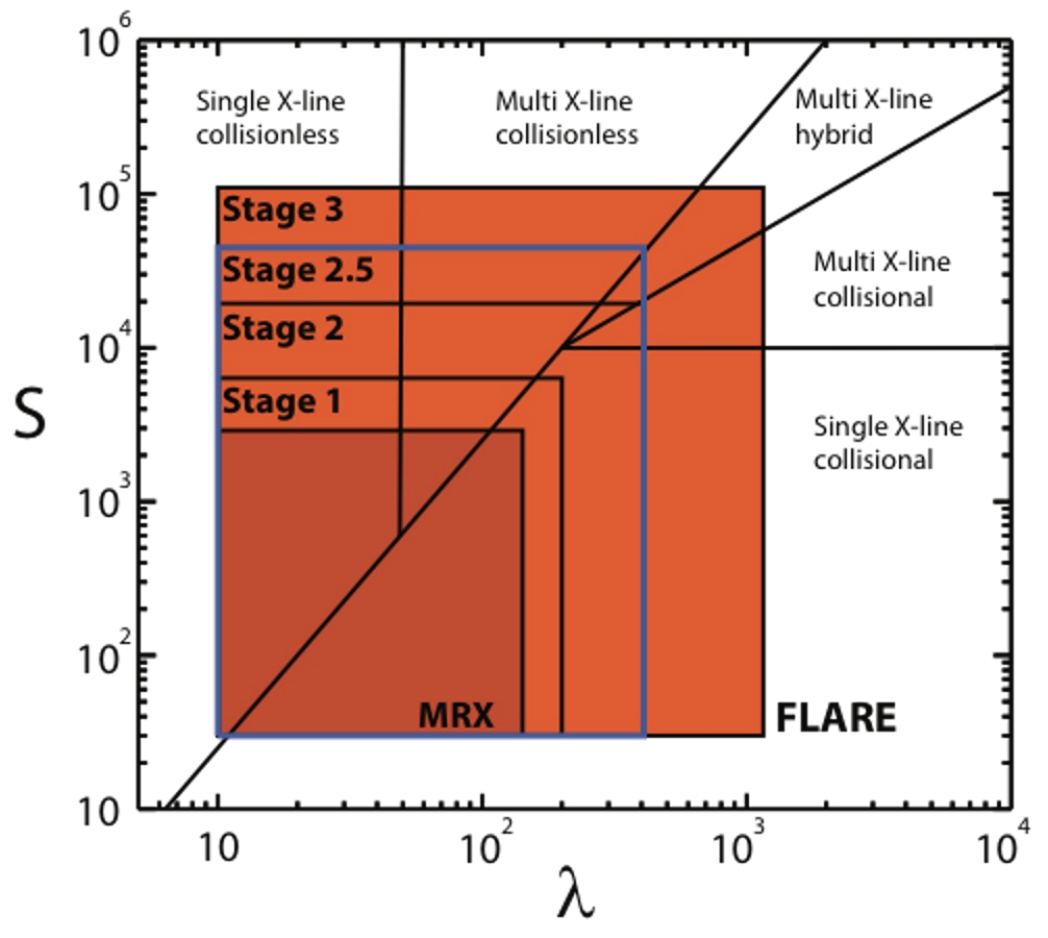}
  \caption{Different stages for the FLARE device to reach different parameter spaces in the phase diagram. \textit{Stage 2.5} is currently available for initial operations and research.}
   \label{fig:stages}
\end{figure}

Experimental capabilities are characterized by stages in the reconnection phase diagram depending on availability and power level of capacitor bank systems (Fig.~\ref{fig:stages}). The previous first plasma operation at Princeton University  achieved  \lq\lq Stage 1" capabilities as designed, providing access to reconnection in a larger parameter space than  MRX. At \lq\lq Stage 2", the first access to multiple X-line collisional and hybrid phases becomes possible while scaling studies with $S$ and $\lambda$ will be enabled in the multiple X-line collisionless phase. At \lq\lq Stage 3", scalings of these new reconnection phases over $S$ or $\lambda$ become feasible. Scalings with $S$ and $\lambda$ are important in order to both extrapolate results to space and astrophysical reconnection cases, as well as to compare with numerical results. The required power levels to achieve each stage are listed in Table \ref{t:banks}. 

All subsystems are currently available except for: two pairs of drive coils, which reduces the reconnection drive, and thus achievable $S$; and structural bracing of GF coils, which limits the GF coil current to those supplied only by one GF bank, and thus achievable $\lambda$. The predicted parameter space is shown as \textit{Stage 2.5} in the reconnection phase diagram (Fig.~\ref{fig:stages}) and in Table~\ref{t:banks}. One particular feature of \textit{Stage 2.5} is that a wide parameter space is available for scaling studies deep in the multiple X-line collisionless phase, in which many space and astrophysical reconnection phenomena occur. This will be our primary focus topic area during the initial operation and research.

\begin{table}
\begin{center}
\begin{tabular}{llllllllllll}
\multicolumn{2}{l}{Capacitor Bank} & OH & EF & GF-A,B,C & PF-A & PF-B & TF-A & TF-B & DC-I & DC-O & Total\\
\hline
\multicolumn{2}{l}{Voltage (kV)} & 20 & 1.4 & 20 & 20 & 20 & 20 & 20 & 20(9)$^*$ & 20(9)$^*$ \\
\hline
\multirow{3}{*}{Stage 3}
& Current (kA) & 180 & 26 & 40 & 540 & 540 & 250 & 250 & 80 & 75 \\
& Energy (MJ) & 0.58 & 0.44 & 3.84 & 0.53 & 0.53 & 0.16 & 0.16 & 0.032 & 0.032 & 6.30\\
& \# of Caps & 6 & 8 & 30 & 6 & 6 & 2 & 2 & 2 & 2 & 64\\
\hline
\multirow{3}{*}{Stage 2.5}
& Current \% & 100\% & 100\% & 33\%$^{**}$ & 100\% & 100\% & 100\% & 100\% & 0\% & 0\%\\ 
& Energy (MJ) & 0.58 & 0.44 & 0.43 & 0.53 & 0.53 & 0.16 & 0.16 & 0 & 0 & 2.82\\
& \# of Caps & 6 & 8 & 10 & 6 & 6 & 2 & 2 & 0 & 0 & 40\\
\hline
\multirow{3}{*}{Stage 2}
& Current \% & 0\% & 100\% & 33\%$^{**}$ & 100\% & 0\% & 100\% & 0\% & 0\% & 0\%\\ 
& Energy (MJ) & 0 & 0.44 & 0.43 & 0.53 & 0 & 0.16 & & 0 & 0 & 1.55\\
& \# of Caps & 0 & 8 & 10 & 6 & 0 & 2 & 0 & 0 & 0 & 26\\
\hline
\multirow{3}{*}{Stage 1}
& Current \% & 0\% & 50\% & 15\% & 58\% & 0\% & 100\% & 0\% & 0\% & 0\%\\ 
& Energy (MJ) & 0 & 0.11 & 0.26 & 0.18 & 0 & 0.16 & & 0 & 0 & 0.70\\
& \# of Caps & 0 & 2 & 2 & 2 & 0 & 2 & 0 & 0 & 0 & 8\\
\end{tabular}
\caption{Maximum current for each power supply at each stage with corresponding stored energy and numbers of capacitors. The GF bank consists of three sub-banks: GF-A, GF-B, and GF-C. (OH=Ohmic Heating, EF=Equilibrium Field, GF=Guide Field, PF=Poloidal Field, TF=Toroidal Field, DC=Drive Coil.) $^*$For DCI and DCO coils, the capacitor bank voltage is limited to 9 kV (out of 20 kV) for the listed maximum currents. $^{**}$The GF coil current is limited by the maximum allowable crowbar diode current of 14 kA at the capacitor bank voltage of 11.5 kV (out of 20 kV). \label{t:banks}}
\end{center}
\end{table}%

FLARE employs 11 capacitor banks to power its internal and external coil systems. The toroidal-field (TF) coils are driven by two identical banks, TFA and TFB, while the poloidal-field (PF) coils are supplied by two identical banks, PFA and PFB. The drive-coil system uses two additional banks --DCI and DCO-- which independently power the inner and outer drive coils, respectively. The Ohmic-heating and equilibrium-field systems each have their own dedicated banks, labeled OH and EF. Finally, the guide-field (GF) coils are powered by three identical banks, GFA, GFB, and GFC, allowing flexible amplitude control, improved balancing, and higher total current capability.

Overall, the system contains five unique capacitor-bank types. The TFA, TFB, DCI, and DCO banks share the same electrical design and component specifications, while the PFA and PFB banks form another pair of identical units optimized for PF-coil operation. The remaining banks—OH, EF, and the three GF banks—each follow designs tailored to their specific coil requirements. This modular arrangement provides the flexibility needed to energize multiple coil subsystems with independent timing, current amplitudes, and pulse shapes, enabling precise control of FLARE’s magnetic configuration during reconnection experiments.

\subsubsection{TF/DCI/DCO/PF/OH banks}

\begin{figure}
   \begin{center}
   \includegraphics[width=1\linewidth]{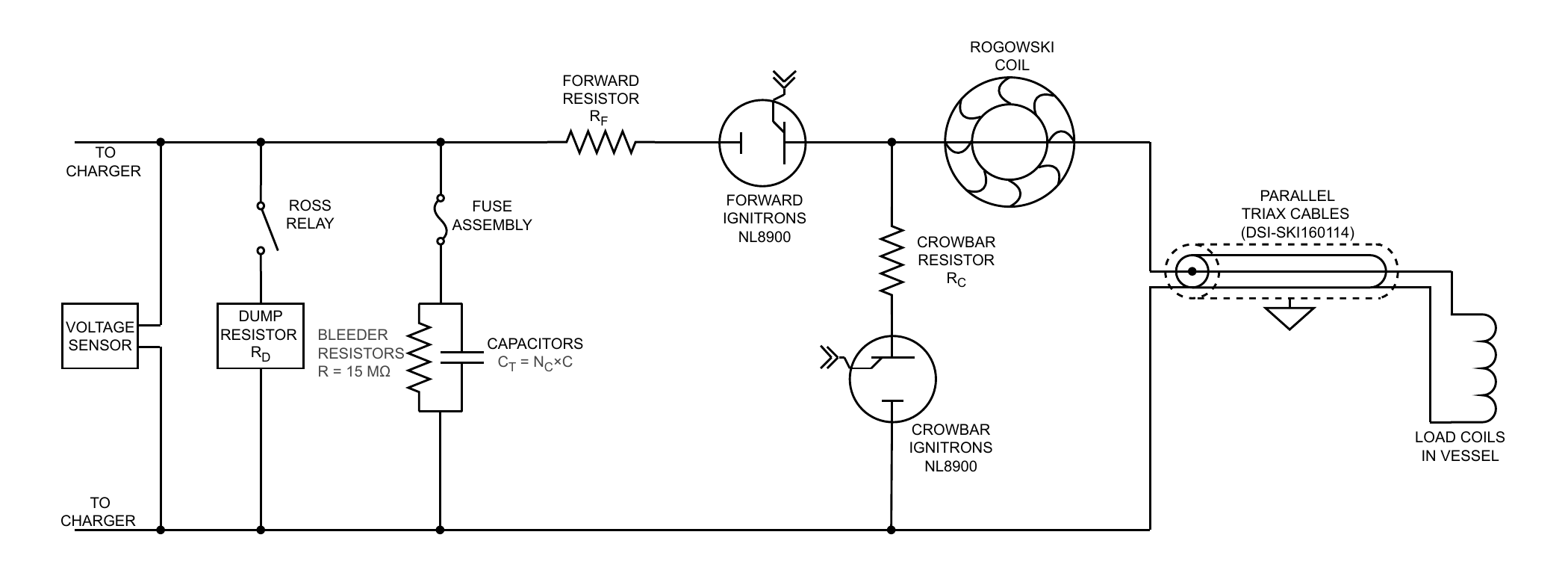}
   \caption{Common circuit schematic used by seven of the FLARE capacitor banks. Each bank consists of a capacitor array with total capacitance $C_{\mathrm{T}} = N_{\mathrm{C}} \times C$, charged by two chargers and protected by individual fuse assemblies and $15~\mathrm{M}\Omega$ bleeder resistors. The discharge current is switched by forward NL8900 ignitrons, passes through a forward resistor $R_{\mathrm{F}}$, and is then delivered to the in-vessel load coils through parallel triaxial cables. A crowbar branch, consisting of crowbar ignitrons and a crowbar resistor $R_{\mathrm{C}}$, is triggered near the peak load current to prevent capacitor voltage reversal and dissipate stored energy. A dump resistor $R_{\mathrm{D}}$ provides controlled energy removal during an aborted discharge. Bank voltage and load current are monitored by voltage sensors and a Rogowski coil, respectively.}
   \label{fig:cb}
   \end{center}
\end{figure}

Among the 11 capacitor banks used in FLARE, seven banks share the same circuit schematic shown in Fig.~\ref{fig:cb}. Differences in component values and the number of components for each bank type are summarized in Table~\ref{t:ccdps}. Each bank is equipped with two charging supplies that connect to the capacitor array during the charging phase, and all banks have a designed maximum charging voltage of $V_{\text{max}} = 20~\mathrm{kV}$. The number of capacitors ($N_{\text{C}}$) and the total capacitance of each bank are chosen to meet the required rise time and peak coil current for the corresponding coil system. A forward resistor $R_{\text{F}} = 1~\mathrm{m}\Omega$ is installed to limit the current in the event of an accidental short on the load side, and each capacitor includes a dedicated fuse assembly for self-protection under fault conditions.

Each bank incorporates a crowbar circuit, which is normally triggered near the time of maximum load current to prevent voltage reversal across the capacitors. A crowbar resistor $R_{\text{C}}$ limits the crowbar current and dissipates most of the stored energy; its value is selected such that the temperature rise during a worst-case discharge remains below 180~K. In addition, each bank includes a dump circuit that can divert energy when a discharge must be aborted. Every capacitor is also equipped with a 15~M$\Omega$ bleeder resistor to safely dissipate stored energy over long timescales in case of system malfunction.

High-voltage, high-current switching is performed using NL8900 ignitrons. The number of ignitrons in both the forward and crowbar stages is determined by the peak-current requirement ($I_{\text{max}}$) for each bank, as summarized in Table~\ref{t:ccdps}. The actual maximum current delivered to a coil is typically lower than $I_{\text{max}}$, due to stray inductance and limits imposed by the coil design.

The capacitor voltage is monitored by two redundant voltage sensors, and the load current is measured using a Rogowski coil. These diagnostic signals are used for both machine protection and operational verification. Power is delivered from each bank to the coils through multiple triaxial cables, with the number of cables ($N_{\text{TC}}$) selected based on the maximum expected current and the required rise time. Increasing $N_{\text{TC}}$ reduces both the stray inductance and resistance of the cable bundle, thereby increasing the achievable load current and decreasing the current rise time.

\begin{table}
\begin{center}
\begin{tabular}{p{10mm}P{12mm}P{12mm}P{6mm}P{12mm}P{12mm}P{12mm}P{6mm}P{6mm}P{6mm}}
Bank  & $I_{\text{max}} (kA) $  & $C_{\text{T}}$ (mF)  & $N_{\text{C}}$  & $R_{\text{F}}$ (m$\Omega$)  & $R_{\text{C}}$ (m$\Omega$) & $R_{\text{D}}$ (k$\Omega$) & $N_{\text{FI}}$  & $N_{\text{CI}}$ & $N_{TC}$ \\ \hline

TF & 250 & 0.8 & 2 & 1 & 3.3 & 3.6 & 2 & 2 & 6 \\ \hline
DCI &  250 & 0.8 & 2 & 1 & 3.3 & 3.6 & 2 & 2 & 6 \\ \hline
DCO &  250 & 0.8 & 2 & 1 & 3.3 & 3.6 & 2 & 2 & 3 \\ \hline
PF &  540 & 2.64 & 6 & 1 & 7.6 & 0.9 & 3 & 3 & 10 \\ \hline
OH &  180 & 2.91 & 6 & 1 & 7.6 & 0.9 & 3 & 3 & 6 \\ 

\end{tabular}
\end{center}
  \caption{Key design parameters of the five FLARE capacitor-bank types that use the common circuit topology shown in Fig.~\ref{fig:cb}. Listed are the maximum design current $I_{\text{max}}$, total capacitance $C_{\text{T}}$, number of capacitor assemblies $N_{\text{C}}$, forward, crowbar, and dump resistances ($R_{\text{F}}$, $R_{\text{C}}$, and $R_{\text{D}}$), the numbers of forward and crowbar ignitrons ($N_{\text{FI}}$ and $N_{\text{CI}}$), and the number of triaxial cables $N_{\text{TC}}$. The actual peak current delivered to a given coil is typically lower than $I_{\text{max}}$ because of stray inductance and limits imposed by the coil system.} \label{t:ccdps}
\end{table}%

\subsubsection{GF banks}

\begin{figure}
   \begin{center}
   \includegraphics[width=1\linewidth]{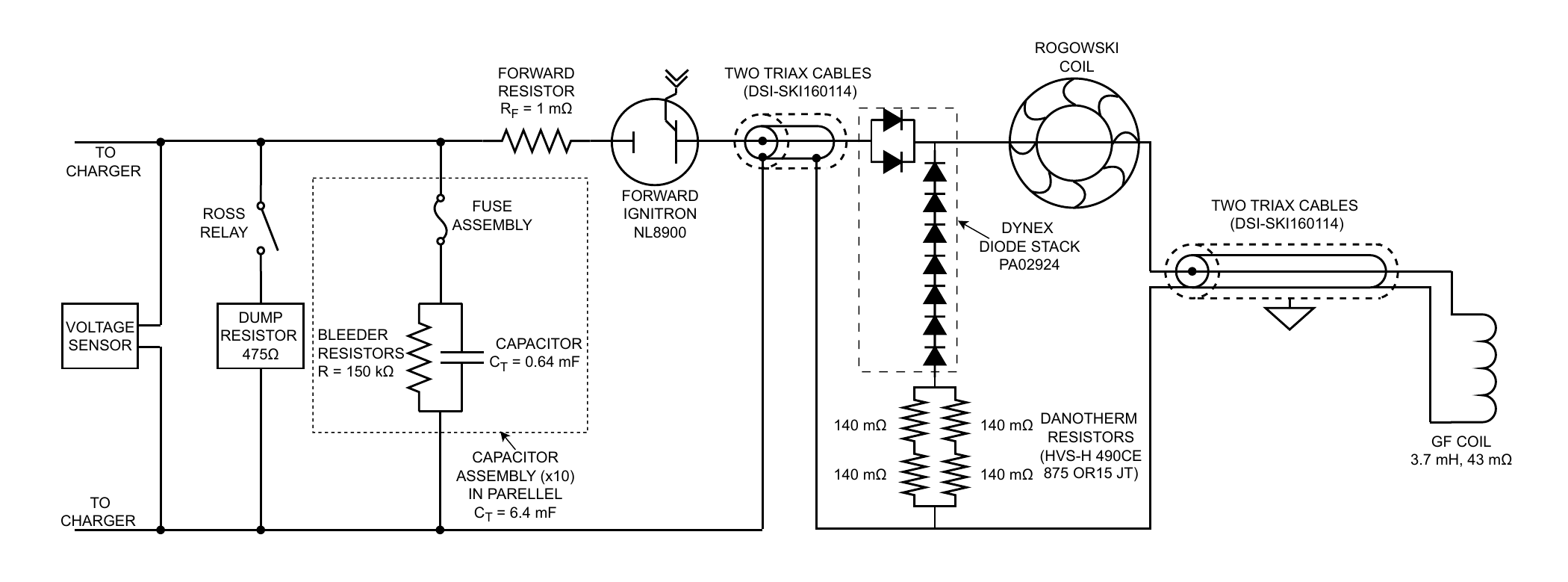}
   \caption{Schematic of the FLARE guide-field (GF) capacitor bank. The bank contains 10 capacitor assemblies in parallel ($C_{\mathrm{T}}=6.4~\mathrm{mF}$) and uses a single NL8900 ignitron as the forward switch. Passive crowbar protection is provided by a Dynex diode stack, and energy is dissipated through Danotherm crowbar resistors and a $475~\Omega$ dump resistor. The circuit also includes fuse and bleeder protection, voltage sensors, a Rogowski coil, and two triaxial cables connected to the GF coil.}
   \label{fig:gfbank}
   \end{center}
\end{figure}

At present, one guide-field (GF) capacitor bank has been completed, with three identical GF banks planned for future operation. Each GF bank follows the circuit schematic shown in Fig.~\ref{fig:gfbank}. Every bank consists of 10 capacitor assemblies connected in parallel, resulting in a total capacitance of 6.4~mF. The maximum operating voltage is 20~kV, and the maximum current per bank is 14~kA. Owing to the large total capacitance, each bank employs a 150~k$\Omega$ bleeder resistor, which is smaller than above fast banks. Energy dissipation during abortion of a discharge is handled by a dump resistor stack with a total resistance of 475~$\Omega$.

Unlike the other capacitor banks, the GF bank uses a single NL8900 ignitron as the forward switch. On the crowbar side, a stack of Dynex diodes (PA02924) provides passive crowbar protection by automatically conducting when an opposite-polarity voltage develops during the LC oscillation. Power dissipation during the crowbar phase is handled by four Dynotherm resistors. As with all other banks, the GF bank is instrumented with two voltage sensors and a Rogowski coil to monitor capacitor voltage and load current during operation.

\subsubsection{EF bank}

\begin{figure}
   \begin{center}
   \includegraphics[width=1\linewidth]{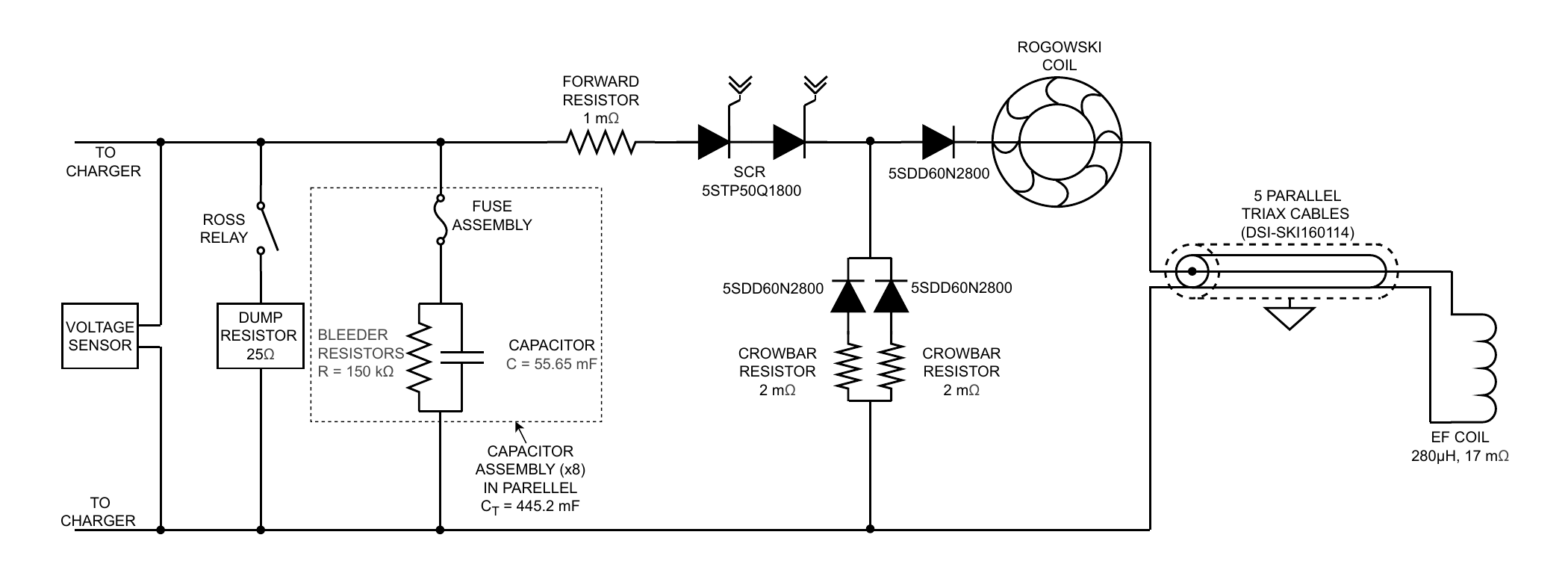}
   \caption{Schematic of the FLARE equilibrium-field (EF) capacitor bank. The bank contains eight capacitor assemblies in parallel ($C_{\mathrm{T}}=445.2~\mathrm{mF}$) and uses two SCRs as the forward switching elements to provide the slow current rise time required for EF operation. Passive crowbar protection is provided by parallel diodes, and energy is dissipated through two crowbar resistors and a $25~\Omega$ dump resistor. The circuit also includes fuse and bleeder protection, voltage sensors, a Rogowski coil, and five parallel triaxial cables connected to the EF coil.}
   \label{fig:efbank}
   \end{center}
\end{figure}

Figure~\ref{fig:efbank} shows the schematic of the equilibrium-field (EF) capacitor bank. To meet the requirement of a slow current rise time, the EF bank is designed with a large total capacitance of 445.2~mF, achieved using 8 capacitors of 55.65~mF connected in parallel. The bleeder resistor has a resistance of 150~k$\Omega$, identical to that used in the GF banks. The maximum operating voltage of the EF bank is 1.4~kV, and the maximum designed current is 26~kA. Energy dissipation during an aborted discharge is handled by a dump resistor stack with a total resistance of 25~$\Omega$.

Unlike other capacitor banks in FLARE, the EF bank uses two silicon-controlled rectifiers (SCRs; 5STP50Q1800) as the forward switching elements. On the crowbar side, two parallel diodes (5SDD60N2800) provide passive crowbar protection by automatically conducting when an opposite-polarity voltage develops during the LC oscillation. Power dissipation during the crowbar phase is handled by two $2~\mathrm{m}\Omega$ crowbar resistors. As with other banks, the EF bank is instrumented with two voltage sensors and a Rogowski coil to monitor the capacitor voltage and load current during operation.

\subsection{Data acquisition}
The FLARE data acquisition system is located inside a RF-shielded screen cage (ETS-Lindgren Model 12, DEI Enclosure). The screen cage provides 100 dB shielding of the electric field and 64 dB shielding of the magnetic field at 14 kHz. Against a plane wave field, it provides over 90 dB shielding up to 1 GHz. Two 1' by 1' waveguide air vents and design with double copper mesh screens provide enough air ventilation for the data acquisition system. 

For digitizing signals, 26 D-TACQ ACQ2106 units with 6 ACQ481 digitizers, which give a total 48 channels per unit operating up to 40 MHz for short captures and 20 MHz for longer captures. Clocking and system-to-system synchronization is provided by two CERN White-Rabbit switches. The input impedance of the digitizer is 100 k$\Omega$. The input voltage range is $\pm 1$ V and it has 14-bit resolution. 

Among 26 units, 25 units (1200 channels) have active integrators. These integrators interface with magnetic probes with the input impedance of 50 $\Omega$. They have controllable gain through programmable gain amplifiers and also have offset control. The frequency range with minimal ($\lesssim 1^{\circ}$) phase shift is from 200 kHz to 5 MHz. The decay time is 1 ms.

\subsection{Control}
The FLARE control system utilizes a distributed architecture built upon the EPICS (Experimental Physics and Industrial Control System) framework, specifically leveraging the pvAccess protocol for high-performance communication. Moving away from legacy LabView-based systems to improve reliability and scalability, the core software stack is developed in Python to simplify the integration of new hardware and diagnostics. The system is designed for open-loop operation with no strict real-time control requirements at the supervisory level; any necessary microsecond-scale responses are implemented locally within hardware, such as FPGAs. A central \textit{Shot Sequencer} manages the experimental cycle through a parallel and asynchronous state machine, which includes distinct phases for Arming, Charging, Firing, Saving data, and Disarming.

The Capacitor Charge/Discharge Power Supplies (CCDPS) control system manages the high-energy (6.2 MJ) capacitor banks that power the FLARE coils. This subsystem is responsible for critical functions such as charge voltage setpoint control, relay sequencing (charge, hold, dump, and ground), and high-speed discharge waveform recording. The design incorporates National Instruments CompactRIO (cRIO) controllers programmed in FPGA mode to handle microsecond-accurate timing and real-time protection logic, such as monitoring ignitron firing status and detecting open-load conditions. Safety is a primary driver of the CCDPS design, featuring an interlocked power source that ensures high-voltage relays fail-safe to a disarm state followed by a grounded state upon power loss or interlock breach.

\begin{figure}
   \begin{center}
   \includegraphics[width=1\linewidth]{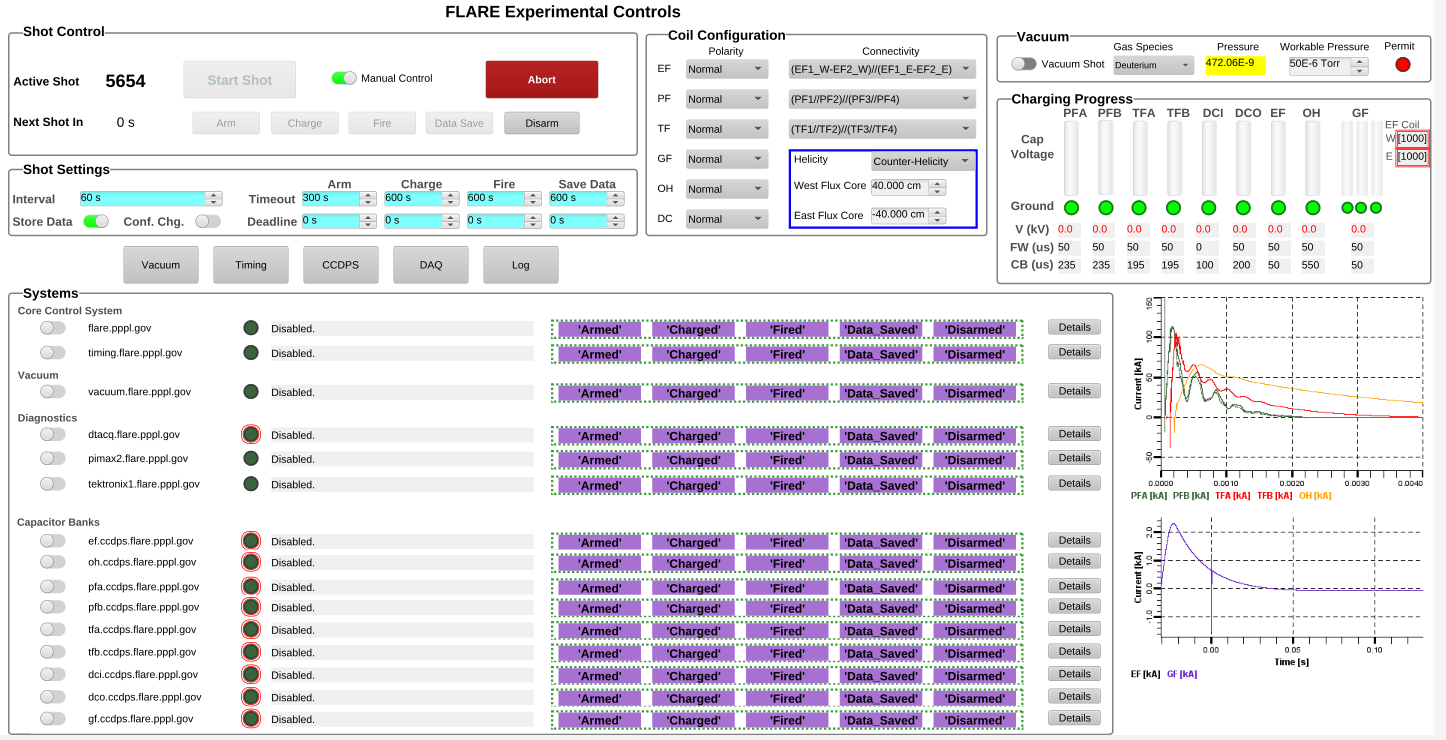}
   \caption{A snapshot of FLARE operation interface based on EPICS.}
   \label{fig:control_opi}
   \end{center}
\end{figure}

The FLARE timing system provides the precision synchronization necessary for experimental operations, utilizing an FPGA-based arbitrary pulse generator capable of 10-nanosecond resolution. It supports up to 32 independent channels for triggering capacitor banks and diagnostics, with all timing parameters recorded in the shot file to ensure experimental reproducibility. Operators interact with the entire facility through unified Graphical User Interfaces (GUIs) developed using Control System Studio (CS:S/Phoebus), as shown in Fig.~\ref{fig:control_opi}. These workstations are isolated on a dedicated FLARE VLAN to enhance cyber security, ensuring that experimental control and data traffic remain separated from the general laboratory network.

Data management for FLARE is centered around a dedicated Data Storage System that utilizes HDF5 as the primary file format for shot data. The HDF5 database allows individual control subsystems to offload experimental data—including raw waveforms and metadata—directly into hierarchical shot files. To ensure data integrity, a multi-file scheme is employed where raw data is stored in a primary file while post-processed analysis results are maintained in separate, linked files. Hierarchical four-level data structure is used for the post-processed analysis results, including level \#1 for raw uncalibrated data as measured by the digitizers, along with calibration coefficients; level \#2 for fully calibrated data in physical units; and level \#3 for data interpolated onto a uniform spatiotemporal grid when appropriate, and for physical variables derived from multiple independent measurements. The hardware infrastructure includes a high-performance server connected via Fiber Channel to the PPPL SAN for long-term archival, supplemented by NVMe-based local caching to provide rapid access to \textit{hot} data for immediate post-shot visualization.

\subsection{Interlock}
The FLARE Interlock System (FIS) is a critical life-safety framework designed to protect personnel from operational hazards by interlocking experimental power sources with emergency-stop (E-Stop) and access control conditions. Developed under the EPICS framework, the system provides a failsafe mechanism to deter entry into hazardous locations and automatically remove hazardous energy if a no access state is breached. The design philosophy prioritizes high reliability and safety integrity, utilizing Nationally Recognized Testing Laboratory (NRTL) listed components and a trapped key methodology to manage access states and ensure procedural compliance during experimental runs.

The system employs a rigorous trapped key sequence to regulate entry into the FLARE test cell and associated high-voltage areas. This methodology ensures that hazardous energy is physically locked out before access keys can be released, and conversely, that all personnel are accounted for and the space is \textit{searched and secured} before the system can be re-energized. To enhance personnel accountability, the final design includes multiple "Diagnostic Access" keys, allowing several individuals to enter the space simultaneously while maintaining individual safety status. Furthermore, the system is designed without a \textit{break glass for key} emergency bypass to prevent intentional or accidental overrides of the safety sequence.

To ensure continuous protection, the FIS incorporates redundant power supplies and an Uninterruptible Power Supply (UPS) for critical control infrastructure. While the redundant power supplies are not individually monitored, the overall UPS status and fiber optic communication links are continuously verified by the system controller. The system logic is programmed to passivate all outputs and generate a system trouble alarm on the Human-Machine Interface (HMI) in the event of a communication loss or remote I/O failure, ensuring the facility defaults to a safe state. Additionally, critical control points such as the rollup door are monitored via trapped key modules to disable motor controllers when the system is in a secure state.

The FIS design strictly adheres to life safety codes, particularly regarding emergency egress. All trapped key-secured doors feature monitored egress buttons that allow personnel to exit the hazardous space in approximately 1-2 seconds, well within the 30-second requirement mandated by laboratory safety standards. The system also integrates a network of E-Stop buttons, which have been upgraded with redundant contacts and clear "FLARE E-STOP" labeling to avoid confusion with other facility systems. These physical safety layers, combined with rigorous Training and Operational Qualification (OQ) plans, form the basis of the independent protection layers (IPL) credited in the project's Layer of Protection Analysis (LOPA).

\subsection{Gas injection}

The FLARE gas injection system is designed to fill the vacuum vessel with a range of working gases, including argon, helium, hydrogen, and deuterium, to a prescribed pressure between 0 and 80~mTorr. To maintain vacuum purity and prevent oil contamination, the gas delivery lines are evacuated using an oil-free mechanical pump (e.g., Rocker~300). The system is rated for a normal operating pressure of 40~psig, a design pressure of 60~psig, and a maximum allowable working pressure of 100~psig.

The primary injection device is a Veeco PV-10 piezoelectric valve. Valve opening is achieved by applying a DC voltage (typically 120~V) from a dedicated gas-injection control circuit that receives a trigger from the FLARE main control system. The control logic employs an open-loop, predictive injection scheme implemented within the EPICS-based framework. In this shot-based approach, a pre-calibrated gas pulse of specified duration and amplitude is delivered to reach the target vessel pressure. The chamber pressure is measured using a Baratron capacitive manometer, which is used for pre-shot vacuum verification and post-shot recording to confirm that the desired gas quantity has been delivered.

Safety considerations are a primary design driver, particularly for the handling of flammable gases such as hydrogen. The system complies with PPPL hydrogen gas codes, requiring all piping to be fabricated from 300-series stainless steel and labeled with standardized yellow tags for clear identification. A dedicated exhaust line is provided for safe venting of the gas-injection plumbing. Burst disks with a threshold pressure of 100~psig are installed to prevent accidental over-pressurization of the gas lines.

To prevent over-pressurization of the vacuum vessel, both the piezoelectric valve and upstream pneumatic valves are interlocked with the vacuum control system. If the chamber pressure exceeds 80~mTorr, the interlock automatically closes the pneumatic valve and disables the gas-injection circuit by removing power through a solid-state relay. In addition, all exhaust from the foreline pump is routed to a dedicated outdoor vent to safely manage any potential hydrogen emissions.

\section{Overview of Initial Diagnostics}

\begin{table}
\centering
\begin{tabular}{p{0.15\linewidth} p{0.35\linewidth} p{0.4\linewidth}}
\textbf{Phase} & \textbf{Diagnostic} & \textbf{Primary Measurements} \\
\hline
\multirow{5}{*}{\textbf{Initial}} 
& Magnetic probe arrays & $\mathbf{B}$ field geometry, current density $\mathbf{J}$, reconnection $E$-field \\
& Triple Langmuir probes & Local $n_e$ and $T_e$ \\
& Fiber-optic interferometer & Line-integrated electron density \\
& Ion Doppler spectroscopy & Ion temperature $T_i$ and line-of-sight flow \\
& Fast camera & Plasma morphology and visible emission \\
\hline
\multirow{9}{*}{\textbf{Additional}} 
& Mach probes & Local ion flow \\
& Fluctuation probes & Local high-frequency fluctuations ($\delta\mathbf{B}$, $\delta\mathbf{E}$, $\delta n_e$) \\
& Dipole probe & Local electric fields \\
& Floating-potential array & 1D floating potential profile \\
& Electron energy analyzer & Electron energy distribution function (EEDF) \\
& Ion Doppler spectroscopy probe (IDSP) & Local ion/neutral temperature and flow \\
& Soft X-ray tomography & 2D profile of energetic electrons \\
& Thomson scattering & 1D profile of $T_e$ and $n_e$, EVDF \\
& Laser-induced fluorescence (LIF) & Neutral temperature and flow velocity \\
\end{tabular}
\caption{Summary of initial and planned additional diagnostics on the FLARE facility. The initial suite focuses on establishing core macroscopic and localized parameters across the extensive operating volume, while the upgrade suite targets advanced kinetic measurements, high-frequency fluctuations, and non-perturbative optical profiling. Optical layouts are shown in Fig.~\ref{fig:ids}(a).}
\label{tab:diagnostics}
\end{table}

The initial diagnostic set on FLARE is designed to provide the core measurements needed to characterize the global evolution of magnetic reconnection over a broad range of plasma conditions. Table \ref{tab:diagnostics} summarizes the phased diagnostic approach of the facility, outlining both the current initial suite and the planned additional diagnostics. In particular, the combined capabilities of the magnetic probe array, Langmuir probes, fiber-optic interferometer, ion Doppler spectroscopy, and fast camera are sufficient to track changes in the global magnetic geometry as functions of the Lundquist number $S$ and the normalized system size $\lambda$. These diagnostics enable direct determination of key reconnection parameters, including the normalized reconnection rate, $S$, and $\lambda$, while also providing measurements of plasma density, magnetic-field structure, and bulk flow. In addition, they allow studies of bulk electron and ion heating and acceleration, and the fast camera provides complementary information on plasma morphology and its time evolution.  

Although this initial diagnostic suite already provides essential measurements of the global and local plasma properties, a more complete understanding of kinetic reconnection requires substantially expanded diagnostic capability. As outlined in the latter half of Table \ref{tab:diagnostics}, FLARE will integrate a number of advanced diagnostics over the next several years to provide a more comprehensive picture of the reconnection layer. For flow, field, and fluctuation measurements, several additional diagnostics are planned. An ion Doppler spectroscopy probe (IDSP) \citep{fiksel98} will measure the local ion and/or neutral temperature and line-of-sight flow velocity. A dipolar probe \citep{stenzel91} will be deployed to measure local electric fields. Mach probes \citep{yoo13,yoo14b} and high-frequency fluctuation probes \citep{ji04,hu21} will be added to measure local plasma flows and electromagnetic and density fluctuations, both of which are critical for understanding anomalous resistivity and wave-particle interactions \citep{yoo24}. For potential and particle-energy measurements, a floating-potential probe array \citep{yoo13,yoo14b} will be used to map global electric fields, and an electron energy analyzer will directly measure the electron energy distribution function and characterize nonthermal particle acceleration. In addition, nonperturbative optical diagnostics, including soft x-ray tomography \citep{xiang21} and Thomson scattering \citep{shi21,shi23}, will provide high-resolution measurements of electron temperature and density profiles, and potentially electron velocity distribution functions \citep{shi22}. Laser-induced fluorescence (LIF) is also planned to measure neutral temperature and velocity \citep{stevenson24}. Figure~\ref{fig:ids}(a) shows their optical layouts. Together, this phased diagnostic suite establishes a robust experimental basis for systematic studies of reconnection scaling and dynamics in FLARE.

\subsection{Magnetic probe array}
\begin{figure}
    \centering
    \includegraphics[width=1\linewidth]{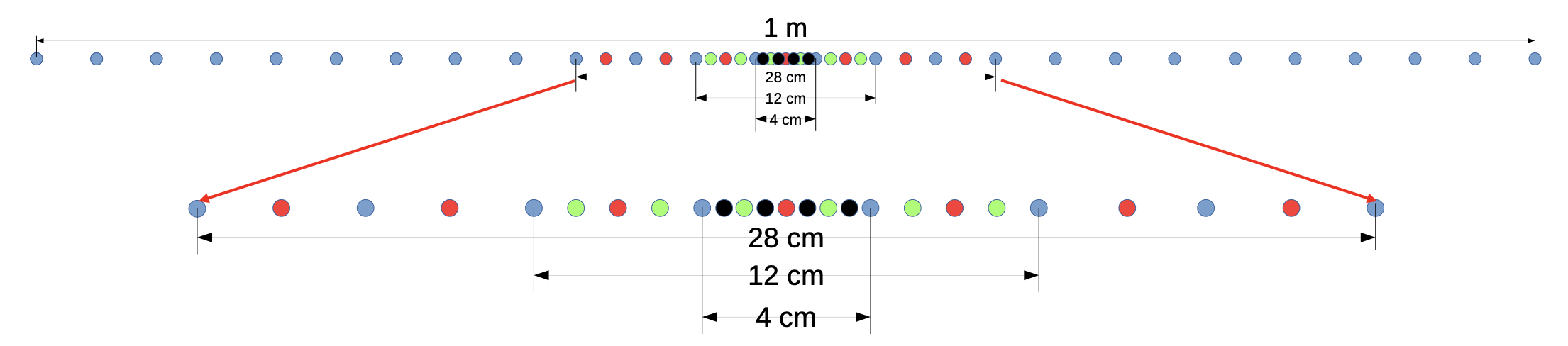}
    \caption{Variable-resolution layout of the FLARE magnetic probe array along the radial direction. Each probe spans a total radial length of $1~\mathrm{m}$ ($R=26$--$126~\mathrm{cm}$) and contains miniature pickup coils for measuring the three magnetic-field components $B_Y$, $B_R$, and $B_Z$. To provide both large spatial coverage and high resolution in the reconnection region, the coil spacing is nonuniform: the central region is instrumented with progressively finer spacing, reaching a maximum radial resolution of $0.5~\mathrm{cm}$ over the innermost $4~\mathrm{cm}$, while wider spacings are used farther from the current sheet, with a base spacing of $4~\mathrm{cm}$ at the outermost positions. The dense central coverage allows accurate reconstruction of sharp gradients in the reconnecting field and current density, whereas the full 1-m extent provides the broad radial coverage required for determination of the global magnetic geometry, poloidal flux, and reconnection electric field. In the initial configuration, 15 such probes are arranged in the $R$--$Z$ plane with a spacing of $3~\mathrm{cm}$ in $Z$.}
    \label{fig:magpb}
\end{figure}
The FLARE magnetic probe array is engineered to simultaneously achieve large spatial coverage and high-resolution measurements of the magnetic field geometry, all while minimizing physical perturbations to the plasma. To withstand the intense thermal fluxes and mechanical stresses of the experimental environment across this extensive length, the probe housing is constructed from custom-made fused quartz ($SiO_2$) tubes. Fused quartz was selected over standard borosilicate glass due to its exceptionally high thermal shock resistance, which mitigates the risk of fracture during high-temperature plasma discharges. The tubes are designed with a 5.1 mm outer diameter and a 3.5 mm inner diameter, utilizing a 0.8 mm wall thickness to guarantee mechanical stability across the probe's long radial extent.

Inside the quartz housing, the probe features an extensive array of 129 miniature magnetic coils configured as triplets to capture the $B_Y$, $B_R$, and $B_Z$ field components. These miniature coils are exceptionally compact, formed as tiny cylinders measuring just 1 mm in height and 1.1 mm in diameter. They are wound with AWG \#44 wire ($14\times4$ turns), yielding an effective area (NA) of approximately $0.2~\text{cm}^2$ and an inductance of $2~\mu$H.

As illustrated in the variable-resolution layout diagram (Fig. \ref{fig:magpb}), the array achieves its dual goal of expansive coverage and precise detail by spanning a full 1-meter radial distance ($R=26$ to $126\text{ cm}$). To optimize data collection across this large coverage area, the coil spacing is variable: within the intense magnetic reconnection region, 75 coils are densely packed over a 28 cm span to achieve a maximum radial resolution of 0.5 cm, whereas the outer extremities utilize a coarser 4 cm base resolution. To ensure precise spatial alignment over this massive distance without twisting, the coils are permanently glued to a custom 1 mm-thick 3D-printed Ultem flat jig prior to their insertion into the quartz tube.

For the initial research phase, the typical configuration will consist of 15 of these magnetic probes arranged to form a comprehensive 2D $R-Z$ plane. The distance between two adjacent probes is set at $3\text{ cm}$ along the $Z$-direction.

With a 2D array of these new magnetic probes, the evolution of all three components of the magnetic field can be measured. The magnetic pickup probe operates on Faraday's law, where a change in the magnetic flux through the pickup coil induces a voltage across the leads of the coil ($V_{\text{coil}}$). The magnitude of the magnetic field component along the axis of the pickup coil is determined by $B = \frac{1}{NA}\int V_{\text{coil}} dt$, where $NA$ is the effective area of the coil. Active analog integrators are used to integrate $V_{\text{coil}}$ before the signals are digitized. These custom integrator circuits employ a second-order multiple feedback bandpass filter topology based on ADA4891 operational amplifiers, which reduces noise, mitigates the impact of digitizer offset voltage, and overcomes bandwidth limitations. The conditioned signals are then routed via customized breakout cables to 8-channel D-tacq digitizers for high-speed recording.The time response of a magnetic probe is limited by the inductance $L$ of the coil and the resistance $R$ seen by the probe. With the inductance of the miniature coil at approximately $2\mu$H and $R$ determined by the $50\Omega$ input impedance of the integrator, the shortest resolvable time scale $\tau$ is $\tau = \frac{L}{R} \approx 40\text{ns}$.This time response is comparable to the digitization rate and fast enough to resolve changes on the Alfv\'en transit timescale ($\sim 1 \mu$s) as well as lower-hybrid waves.

The 2D magnetic field profiles obtained from the array are critical for computing the current density using Ampère's law. By assuming toroidal symmetry—a valid approximation given the inherent global symmetry of the device during quasi-steady periods—each component of the in-plane current density can be reconstructed:
\begin{align}
J_R & = -\frac{1}{\mu_0} \frac{\partial B_\theta}{\partial Z}, \\
J_\theta & = \frac{1}{\mu_0} \left(\frac{\partial B_R}{\partial Z} - \frac{\partial B_Z}{\partial R} \right),\\
J_Z & = \frac{1}{\mu_0 R} \frac{\partial}{\partial R} (RB_\theta).
\end{align}
Furthermore, the array enables the calculation of the out-of-plane reconnection electric field. Assuming axisymmetry, the poloidal flux function $\psi$ is determined as:
\begin{align}
\psi(R,Z,t) & = \int_0^R 2\pi R^{\prime}B_Z(R^{\prime},Z,t)dR^{\prime} \\
& = \int_0^R 2\pi R^{\prime}B_Z(R^{\prime},Z_0,t)dR^{\prime}-\int_{Z_0}^Z 2\pi R B_R(R,Z^{\prime},t)dZ^{\prime},
\end{align}
where $\psi$ is set to zero at the machine axis ($R=0$). Because calculating $\psi$ requires the $B_Z$ profile from the machine center to the measurement region, the expansive 1-meter radial coverage of the new FLARE probes ($R=26$ to $126\text{ cm}$) inherently fulfills this requirement without needing separate, coarser background probes. The inductive out-of-plane reconnection electric field $E_Y$ can then be calculated as $E_Y = -\frac{1}{2\pi R} \frac{\partial \psi}{\partial t}$.

\subsection{Langmuir probe}
\begin{figure}
    \centering
    \includegraphics[width=0.7\linewidth]{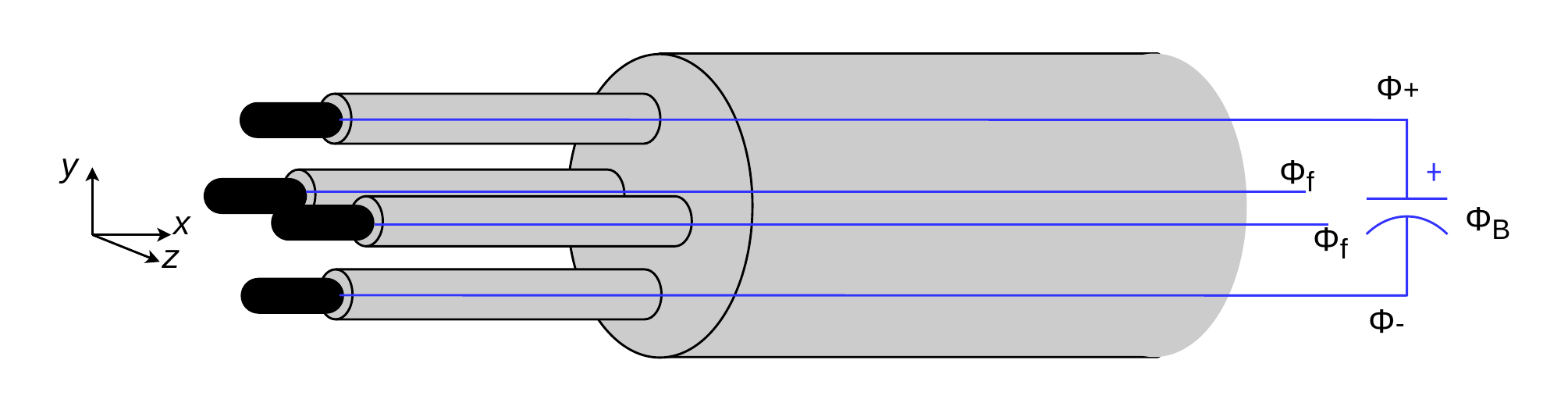}
    \caption{Schematic of the modified triple Langmuir probe configuration used in FLARE. The probe consists of four tungsten tips (0.76 mm diameter, 1.5 mm length) arranged in a 2 mm by 2 mm square geometry. A fixed bias voltage ($\Phi_B$) is applied by a pre-charged capacitor between the electrodes labeled $\Phi_+$ and $\Phi_-$ in a double-probe configuration to drive and collect the ion saturation current. The remaining two unbiased tips provide measurements of the floating potential ($\Phi_f$). By averaging the signals from the two $\Phi_f$ tips, the diagnostic significantly mitigates errors in the electron temperature ($T_e$) calculation induced by the strong, local in-plane electric fields along the outflow ($z$) direction.}
    \label{fig:lang}
\end{figure}

A modified triple probe is used to measure plasma density and electron temperature \citep{chen65,ji91}. A schematic of the probe is given in Fig. \ref{fig:lang} \citep{ji91}. As illustrated, the probe consists of four tungsten tips, each 0.76 mm in diameter and 1.5 mm in length, arranged in a 2 mm by 2 mm square geometry. The probe diagnostic utilizes a double probe configuration for two of the electrodes (labeled $\Phi_+$ and $\Phi_-$), across which a bias voltage ($\Phi_B$) is applied by a series of capacitors charged prior to a discharge, while the remaining two unbiased tips provide parallel floating potential ($\Phi_f$) measurements. Assuming a Maxwellian distribution function, the current-voltage characteristic of the probe is: 
\begin{equation}
  I_p(\Phi) = -I_{sat} + en_eA\sqrt{\frac{T_e}{2\pi m_e}}\exp\left[-\frac{e(\Phi_p-\Phi)}{T_e}\right],    
\end{equation}
where $I_p(\Phi)$ is the current flowing from the probe to the plasma, $A$ is the probe area, and $\Phi_p$ is the plasma potential. The bias voltage $\Phi_B$ is set several times higher than $T_e$, ensuring that the $\Phi_-$ tip collects the full ion saturation current ($I_\text{sat}$). Because the double probe is electrically floating, the $\Phi_+$ tip draws an equal amount of electron current. Using the relations $I_p(\Phi_+) = -I_p(\Phi_-)$ and $I_p(\Phi_f) = 0$, the relationship between $\Phi_+$ and $\Phi_f$ is computed as \citep{chen65,hutchinson05}
\begin{equation}
    \frac{I_p(\Phi_+)-I_p(\Phi_f)}{I_p(\Phi_+)-I_p(\Phi_-)} = \frac{1}{2} = \frac{1-\exp[-(\Phi_+-\Phi_f)/T_e]}{1-\exp(-\Phi_B/T_e)}.
\end{equation}
When $\Phi_B \gg T_e$, this relationship simplifies to $\Phi_+-\Phi_f = (\ln 2)T_e$, allowing for the direct determination of the electron temperature. The primary purpose of utilizing two separate floating potential tips is to mitigate errors induced by local electrostatic electric fields, which can reach up to 1000 V/m and artificially skew $T_e$ measurements by approximately 3 eV over a 2 mm tip separation \citep{yoo13}. Aligning the floating tips along the $z$ direction and averaging the signals from the two $\Phi_f$ tips, with the $\Phi_+$ tip situated between them, significantly minimizes this systematic error \citep{ji91}. 

Plasma density is calculated by measuring the ion saturation current, expressed as $I_{sat} = \exp(-0.5)Aen_iC_s$ where $C_s$ is the ion sound velocity \citep{hutchinson05}. While $C_s$ is typically approximated as $\sqrt{T_e/m_i}$ when $T_e \gg T_i$, the ion temperature in these discharges can be comparable to or slightly larger than the electron temperature. However, provided that $T_i < 2T_e$, the ion saturation current remains relatively insensitive to $T_i$, making $C_s \approx \sqrt{T_e/m_i}$ a valid operational approximation \citep{hutchinson02}. To infer the electron density ($n_e$) from the measured ion density ($n_i$), the effective ionic charge ($Z_\text{eff}$) must be considered. In these discharges, the population of doubly-ionized gas (e.g., helium) is negligible due to high second ionization energies relative to the typical electron temperatures of 5 to 20 eV achieved so far. Coupled with the facility's excellent base pressure capabilities, $Z_\text{eff}$ is expected to remain close to unity. Finally, to account for geometric shielding and sheath expansion effects, the effective probe area $A$ is cross-calibrated using independent density diagnostics (such as laser interferometry). Overall, data derived from this triple Langmuir probe array carries a standard uncertainty of 10\% to 20\%.

\subsection{Fiber-optic interferometer}
\begin{figure}
    \centering
    \includegraphics[width=1\linewidth]{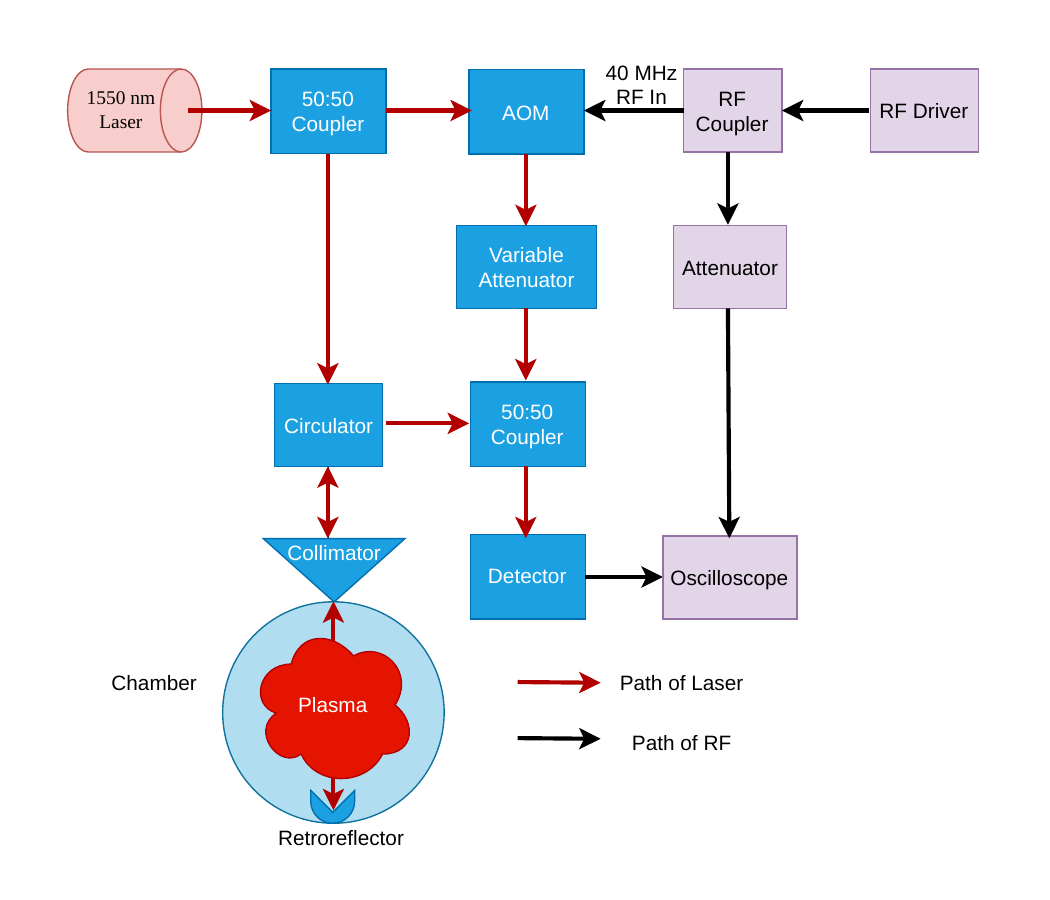}
    \caption{Schematic of the fully fiber-coupled 1550~nm heterodyne Michelson interferometer used on FLARE to measure line-integrated electron density. The output of a distributed-feedback diode laser is split into a reference leg and a plasma-measurement leg by a 50:50 fiber coupler. In the reference leg, an acousto-optic modulator (AOM) driven at 40~MHz imposes a frequency shift, and a variable attenuator is used for intensity matching. In the plasma leg, the beam is launched into free space through a collimator, traverses the FLARE vessel, reflects from an in-vessel retroreflector, and returns through the plasma to the detection system. The returning plasma beam is recombined with the frequency-shifted reference beam at a second 50:50 coupler, producing a 40~MHz beat signal whose phase shift is proportional to the line-integrated electron density. The mixed signal is detected by an amplified photodiode and digitized by an oscilloscope for post-shot phase analysis.}
    \label{fig:interferometer}
\end{figure}
To measure the line-integrated electron density of the plasma, the FLARE diagnostic suite employs a fully fiber-coupled infrared laser heterodyne Michelson interferometer. This design marks a significant operational improvement over traditional free-space systems—such as the 10.6 $\mu\text{m}$ CO2 laser interferometer previously utilized on the Magnetic Reconnection Experiment (MRX), which suffered from vibrational noise and required time-consuming daily realignment. By transitioning to a heavily fiber-coupled topology, the diagnostic achieves a highly robust system that is exceptionally resistant to alignment perturbations and mechanical vibrations.

The core of the system is an Emcore (QDFBLD-1550-100) distributed feedback diode laser, which produces a 100 mW continuous-wave (CW) output. The laser operates at a wavelength of 1550 nm, a standard extensively utilized in the telecommunications industry. Operating at this telecommunications standard allows the diagnostic to leverage a wide variety of readily available, high-quality fiber optic components. Furthermore, the laser provides a very long coherence length, which is critical for reducing phase noise and improving measurement quality given the inherently unmatched lengths of the reference and plasma beam paths.

As illustrated in the system schematic (Fig. \ref{fig:interferometer}), the laser output is split into two distinct paths by a 50:50 fiber coupler. The first path serves as the reference leg, which is routed through an Acousto-Optic Modulator (AOM) driven by a 40 MHz RF signal to induce a precise frequency shift. The shifted reference signal then passes through a variable attenuator to allow for intensity matching. The second path forms the plasma measurement leg. The beam passes through a circulator and is launched into free space via a collimator, resulting in a 7 mm diameter beam. The beam traverses an approximately 6-meter free-space path through the FLARE vacuum chamber, strikes a retroreflector mounted inside the vessel, and returns through the plasma back into the collimator. Because the group velocity of the electromagnetic wave in the plasma differs from its velocity in a vacuum, the beam accumulates a phase shift ($\Delta\phi$) that is directly proportional to the line-integrated electron density along the path.

Upon re-entering the collimator, the return signal is routed by the circulator to a second 50:50 coupler, where it recombines with the 40 MHz frequency-shifted reference beam. This heterodyne recombination generates a 40 MHz beat frequency that carries the accumulated phase shift from the plasma. To prevent detector saturation, the mixed signal is passed through a final variable attenuator before being fed into an InGaAsP amplified photodiode detector. The resulting electrical signal is directly digitized by a high-speed oscilloscope, allowing the phase shift to be computationally extracted during post-shot data analysis.

\subsection{Ion Doppler spectroscopy}

\begin{figure}
    \centering
    \includegraphics[width=1\linewidth]{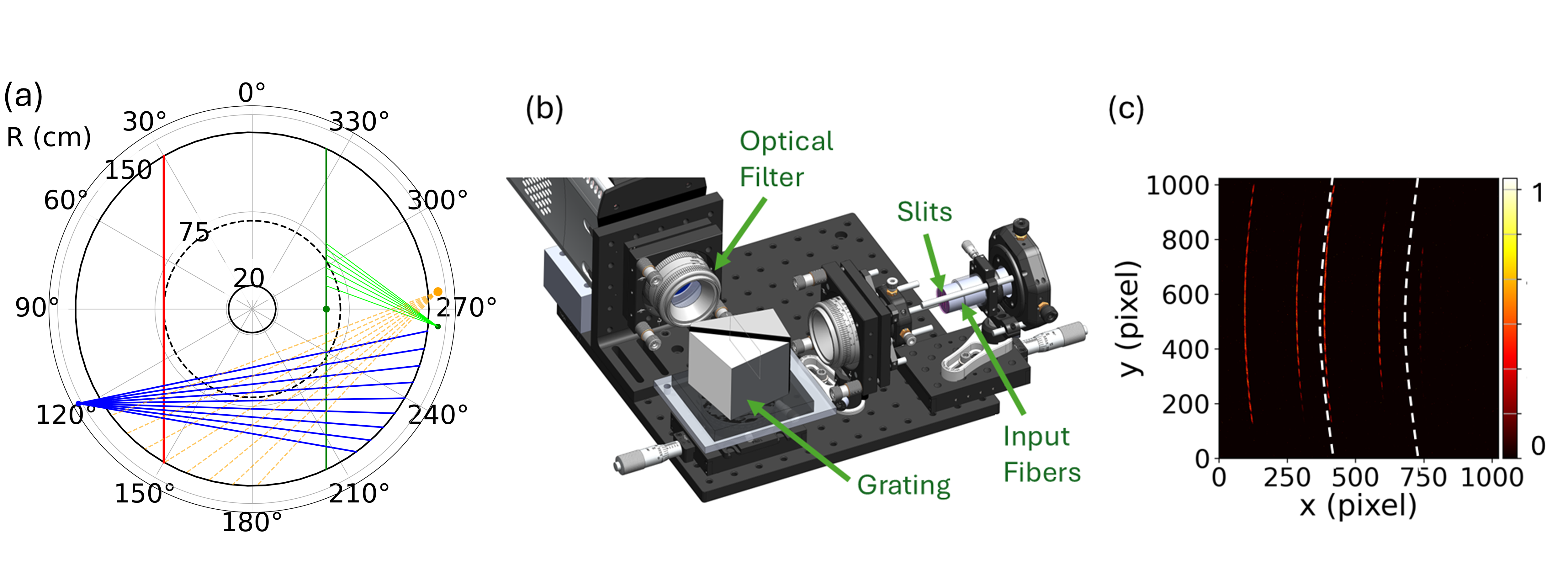}
    \caption{(a) The collection optics layout of ion Doppler spectroscopy (blue lines); vacuum vessel wall, current sheet center and central stack of radius 150 cm, 75 cm and 20 cm are plotted as references. Optics layouts of other diagnostics are also listed, including interferometer (red), soft X-ray tomography (orange) and Thomson scattering/Laser-Induced Fluorescence (green). (b) the high-throughput spectrometer equipped with Volume Phase Holographic (VPH) grating with an F/\#=1.8; (c) Example spectrum obtained during helium plasma discharge. $x$ is the wavelength axis while $y$ stands for different injection fibers. White dashed lines denote bandwidth of the band-pass optical filter.}
    \label{fig:ids}
\end{figure}

The ion Doppler spectroscopy diagnostic has been implemented in FLARE to measure ion temperature and flow speed~\citep{goodman23}. As is shown in Fig.~\ref{fig:ids}(a), the collection optics comprised of 80 optical fibers and bi-convex lenses is used to cover tangency radii from 50 cm to 100 cm, i.e., 25 cm upstream regions of pull-phase reconnection current sheet centered around 75 cm. The spatial resolution is 1 cm, determined by focal length of the collection lenses and core diameter of optical fibers. Since each fiber collects the integrated plasma emission light along the light of sight, the multiple-collection-path design enables tomography inversion to reconstruct the radial profiles of ion temperature and drift speed, under the assumption that plasma processes are axisymmetric~\citep{bell21}.

To achieve a large optic throughput while maintaining a high wavelength resolution, a spectrometer equipped with volume phase holographic grating of an F/\#=1.8 is designed and developed as shown in Fig.~\ref{fig:ids}(b). The direct consequence of the chosen short focal length 50 mm is that spectra of light from off-axis fibers would be curved~\citep{bell04}, as presented in Fig.~\ref{fig:ids}(c). The other novel feature for this design is to arrange three columns of fibers separated horizontally (along the wavelength axis $x$) so that spectral lines of same wavelength from different columns will be imaged at different $x$ pixels on the iCCD (Intensified Charge-Coupled Device). In this way, this spectrometer allows for more input fibers and further enables spectral measurements from more collection channels. To avoid spectral image overlapping from different columns, the band-pass optical filter of [467.5 nm, 471.6 nm] is inserted before the imaging lenses of iCCD, denoted by white dashed lines in Fig.~\ref{fig:ids}(c). Moreover, the optical slits of width 40 $\mu\text{m}$ are placed immediately in front of injection fibers to ensure the instrumental spectral width around 0.04 nm, corresponding to helium ion temperature of 5 eV.

\subsection{Fast camera}
\begin{figure}
    \centering
    \includegraphics[width=1\linewidth]{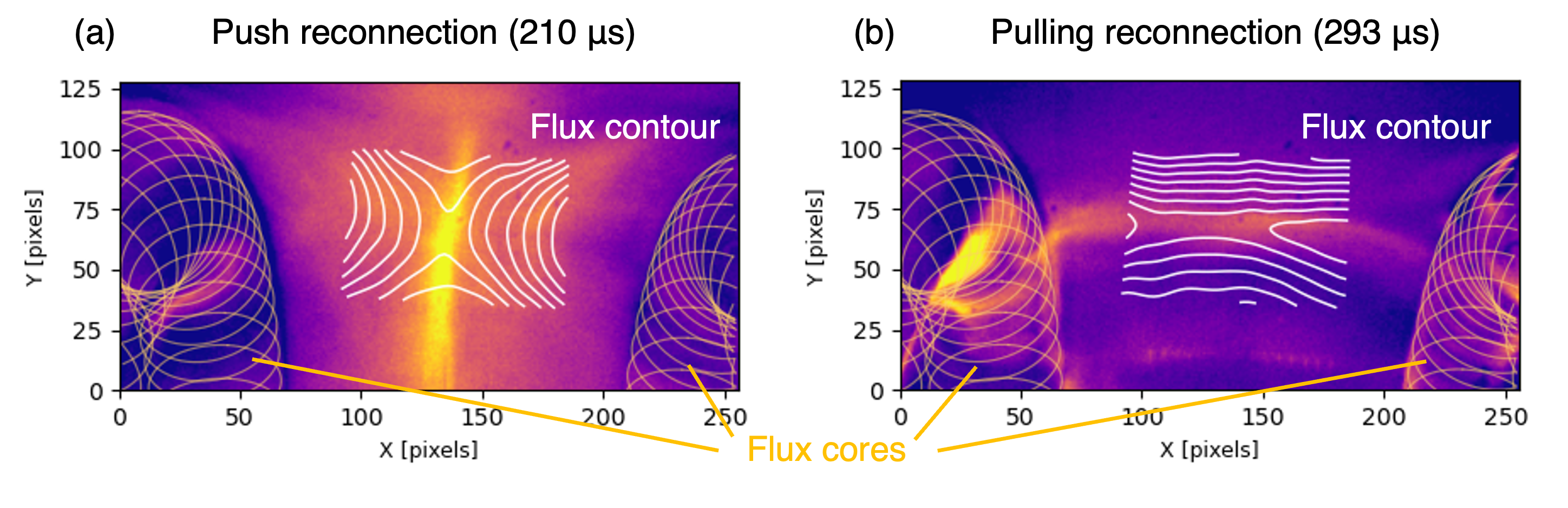}
    \caption{Fast camera images during (a) pushing and (b) pulling reconnection with an exposure time of 1~$\mu$s. The images are processed by subtracting the average of seven frames around the target frame, and the color intensity is normalized. The 3D flux core structures are overlaid as orange solid lines to demonstrate the quality of the camera calibration, and 2D flux contours reconstructed from the magnetic probe array are shown as white solid lines.}
    \label{fig:fastcamera}
\end{figure}

The FLARE device is equipped with a fast camera diagnostic to qualitatively visualize plasma emission during the discharge, as shown in Fig.~\ref{fig:fastcamera}. The camera (Phantom v7.3, CMOS sensor) is installed with a tangential viewing geometry, oriented upward toward the magnetic probe array, and uses a fisheye lens to maximize the field of view. It operates at a frame rate of 60,000 frames per second ($\sim$16~$\mu$s) for a wider field of view (256 $\times$ 128 pixels), corresponding to a typical local scale of $\sim$4.7~mm/pixel. For higher temporal resolution, the camera can be operated at up to 120,000 frames per second ($\sim$8~$\mu$s) with a reduced field of view (128 $\times$ 64 pixels). The exposure time is 1~$\mu$s. No optical filter is used, and the camera records all visible emission, most likely dominated by H$\alpha$ emission in deuterium plasma. A fisheye lens is employed to maximize the field of view. The resulting geometric distortion is corrected through a calibration procedure based on matching known vessel structures (e.g., flux cores), which are indicated by the orange solid lines in Fig.~\ref{fig:fastcamera}.

To facilitate qualitative comparison between diagnostics, magnetic flux contours reconstructed from the magnetic probe array are overlaid on the fast camera images during both the pushing reconnection phase (Fig.~\ref{fig:fastcamera}(a)) and the pulling reconnection phase (Fig.~\ref{fig:fastcamera}(b)).This shows good agreement between the observed emission structures and the inferred magnetic topology from the magnetic probe array. The images here are processed by subtracting a background image constructed from the average of seven frames around the target frame ($\sim$112~$\mu$s), which enhances rapidly evolving plasma structures.

\section{Initial Operations Capabilities}
FLARE has successfully entered its operational phase, demonstrating the mechanical and electrical capability to access an unprecedented parameter space for laboratory magnetic reconnection studies. The primary engineering objective of FLARE is to bridge the gap between single–X-line reconnection regimes and multiple–X-line regimes relevant to astrophysical, heliospheric, and fusion plasmas. As shown in Fig.~\ref{f:phase}, this transition can be characterized by two key dimensionless parameters, $S$ and $\lambda$. In its current operational state within \textit{Stage 2.5}, the machine routinely produces plasmas with $S \sim 2{,}500$ and $\lambda \sim 60$ for anti-parallel reconnection. The relevant parameters are: electron density $n = 0.5$--$30\times 10^{19}$ m$^{-3}$, electron temperature $T_e = 10$--$15$ eV, reconnecting field component $B_{rec} \sim 0.05$ T, and system size $L = 0.8$ m.

With the planned increased energy per discharge, FLARE is projected to significantly extend its accessible parameter space. This parameter space is highly tunable due to the machine’s flexible mechanical and electrical design. For example, the separation between the two primary flux cores can be adjusted between 0.8~m and 1.6~m, enabling direct control of the macroscopic system size. The modular capacitor-bank system provides independent control of the poloidal-field (PF) and toroidal-field (TF) coils of each flux core. In addition, guide-field (GF) and Ohmic-heating (OH) coils are independently controlled, allowing fine shot-to-shot tuning of the initial magnetic topology, plasma heating and reconnection drive.

\subsection{Available Operational Modes on FLARE\label{s:mode}} 

\begin{figure}
  \centering
  \includegraphics[width=5in]{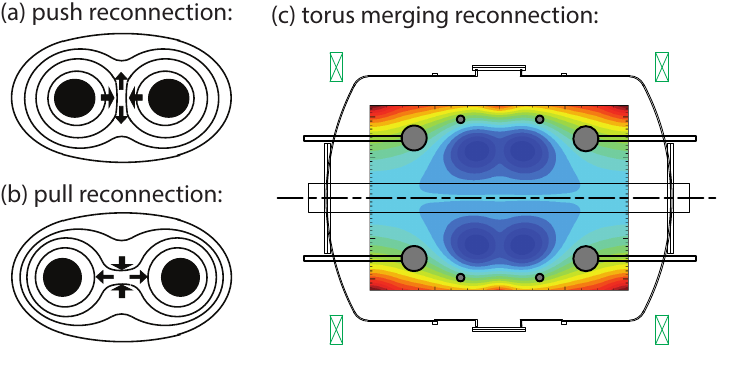}
  \caption{(a) \lq\lq push" reconnection and (b) \lq\lq pull" reconnection driven by flux cores (in black), and (c) reconnection during merging of toroidal plasmas after they are \lq\lq pinched off" or detached from the flux cores. The color plot is contours of constant poloidal flux from a 2D MHD simulation of counter-helicity spheromak merging adapted from \cite{yamada07b}.}
   \label{fig:push-pull-merging}
\end{figure}

\underline{\textit{\lq\lq Push" versus \lq\lq pull" versus merging reconnection}}. Although the primary mode of FLARE operation is the so-called \lq\lq pull" reconnection mode as shown in Fig.~\ref{f:MRX}, there are two additional operational modes to drive reconnection using flux cores as illustrated in Fig.~\ref{fig:push-pull-merging}. The \lq\lq push" reconnection mode is induced by further increasing current in the PF coils of the flux cores. Push reconnection has been used in MRX to study the Hall effect during collisionless reconnection~\citep{inomoto06} depending on the sign of radial component of reconnection current or electric field: Case I for positive sign while Case O for negative sign. On FLARE, however, it is potentially more useful because (1) the time period is expected to be longer due to overall larger inductance, and (2) plasma production can be made much earlier by the OH coils, especially with strong guide field, so that the push phase of reconnection can start much earlier and therefore last longer.

Another way to drive reconnection is to merge two plasma toroids after they detach from the flux cores by delaying the crowbar switch in the PF circuit. These toroidal plasmas can be spheromaks~\citep{yamada81} or spherical tokamaks~\citep{tanabe17} depending on absence or presence of guide (toroidal) field. By creating and merging two of these plasmas, controlled study of magnetic reconnection and subsequent plasma energization can be performed~\citep{yamada90,ono96,gerhardt07,ono15}.

Two additional methods for driving pull reconnection are available on FLARE. The first one is by two or more pairs of drive coils. This is effective in enhancing pull reconnection as shown in MRX~\citep{jara-almonte16}. The second method is by using ohmic heating coils, which further drive pull reconnection, in addition to heating the plasma. This method has not been tested on MRX but is similar to ohmically-driven fusion experiments. Initial kinetic simulations have shown that OH drive may efficiently form plasmoids.

\underline{\textit{Anti-parallel versus guide field reconnection}}.
In addition to the in-plane (poloidal) component, the out-of-the-plane (toroidal) component along the X-line, often called guide field, can be adjusted from nearly zero (anti-parallel reconnection), as occurs in Earth's magnetotail, to orders of magnitude larger than the reconnecting component, as in solar reconnection or fusion plasmas. Compared to MRX where the ratio of guide-to-reconnecting field can be varied from zero to only a few~\citep{tharp12,fox17,bose24}, the ratio on FLARE is increased up to $\sim100$. 

\underline{\textit{Co-helicity versus counter-helicity reconnection}}.
During push and merging reconnection, there is additional freedom in the direction of the current in the TF circuit. If the toroidal field is chosen to have the same sign for each flux core, the magnetic helicity of each plasma will have also the same sign, leading to co-helicity reconnection. The opposite choice leads to counter-helicity reconnection~\citep{yamada90}. Locally, counter-helicity reconnection is equivalent to anti-parallel reconnection while the co-helicity configuration corresponds to guide field reconnection. Globally, however, co-helicity reconnection differs from guide field reconnection in that the toroidal component is generated by currents flowing within the plasma as opposed to flowing within the guide field coil. Both choices of plasma merging process are being used either in spherical tokamak startup or in Field Reversed Configuration (FRC) devices.

\underline{\textit{Symmetry and boundary conditions}}.
In its idealized form, reconnection is highly symmetric with mirror symmetry in both upstream and downstream. These symmetries are often broken in real applications, such as in magnetopause reconnection and fusion plasmas where asymmetry arises in upstream, or in magnetotail reconnection and solar reconnection where asymmetry arises in downstream. As in MRX~\citep{yoo14}, controlled upstream asymmetries can be imposed on FLARE, enabling systematic studies of asymmetric reconnection which can be important for the suppression and onset of reconnection~\citep{zakharov92,swisdak03,phan10} in much larger parameter ranges. In addition, downstream asymmetry can be studied on FLARE since TF and PF coils in each flux core is powered independently. The toroidal symmetry can be also broken by \textit{e.g.}, placing electrically conducting plates at certain toroidal angles which introduce line-tied boundary conditions. On MRX, this has been not tried in the configurations shown in Fig.~\ref{fig:push-pull-merging}, but has been used in experiments to study flux rope stability for the solar applications~\citep{myers15}. FLARE will offer flexibility and controls of both symmetry and boundary conditions.

\underline{\textit{Plasma properties and composition}}. FLARE will offer a wide range of ionization fraction (from full ionization, partial ionization, to poorly ionization as demonstrated~\citep{lawrence13} previously on MRX) by controlling the gas fill pressure and/or coil currents. Ion species and compositions can be selected by using a combination of diatomic molecular gases (such as H$_2$, D$_2$, N$_2$), monatomic noble gases (such as He, Ne, Ar, and Kr), and other gases (such as methane CH$_4$). We also plan to accommodate special requests from users who may bring their own gases, such as the exotic $^3$He gas.

\subsection{Operation Mode Examples}

\begin{figure}
    \centering
    \includegraphics[width=1\linewidth]{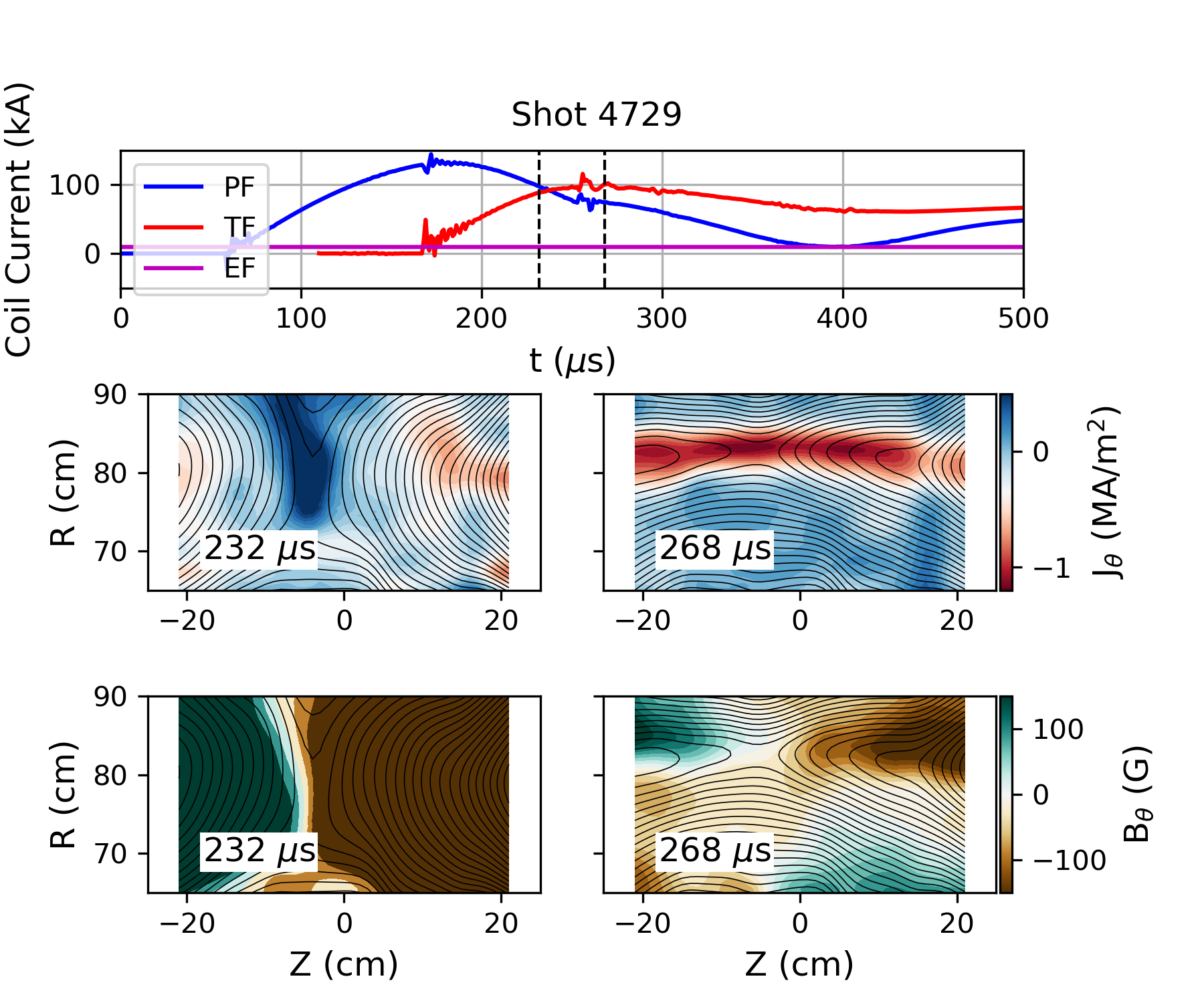}
    \caption{Typical push-pull reconnection discharge in FLARE with deuterium in counter-helicity, Case-I operation. The top panel shows the PF, TF, and EF coil currents. The middle panels present the out-of-plane current density $J_{\theta}$, and the bottom panels show the out-of-plane magnetic field $B_{\theta}$, at $t=232~\mu\mathrm{s}$ for push reconnection and $268~\mu\mathrm{s}$ for pull reconnection. Black lines are poloidal flux contours.}
    \label{fig:pushpull}
\end{figure}

The flexibility of FLARE enables access to a variety of reconnection configurations through independent control of the PF, TF, EF, GF, and OH coil systems. In particular, the coils associated with the two flux cores can be driven independently, allowing controlled asymmetry in both upstream and downstream boundary conditions. In this subsection, we present three representative operation modes that demonstrate the capability of FLARE to realize distinct reconnection geometries and dynamics. All examples shown here correspond to counter-helicity, Case-I operation \citep{inomoto06}. In addition, although the GF coil is not activated in these examples, it can be used to increase the normalized system size, improve plasma stability, and modify the overall reconnection dynamics \citep{tharp12,fox17,bose24}.

The first example, shown in Fig.~\ref{fig:pushpull}, corresponds to a push-pull configuration using deuterium as the working gas. The top panel shows the coil currents for the PF, TF, and EF systems. In this discharge, the waveforms of PFA and PFB are nearly identical, and those of TFA and TFB are also nearly identical, indicating a highly symmetric operating condition. 

The middle panels show contours of the out-of-plane current density, $-J_y$, while the bottom panels show the out-of-plane magnetic field component, $B_y$. At $t = 232~\mu$s, a push current sheet is formed as the two plasmas are driven toward each other by the plasma pressure. The out-of-plane magnetic field generated during plasma formation \citep{yamada95,yamada97a,yamada97b} is already present at this time: the left side has positive $B_{\theta}$, whereas the right side has negative $B_{\theta}$, consistent with the expected polarity for Case-I operation. The push current sheet exhibits positive $J_{\theta}$ and is elongated primarily along the $R$ direction. At this stage, the large preexisting $B_{\theta}$ component associated with plasma formation obscures any clear Hall-like quadrupole structure. 

At a later time, $t = 268~\mu$s, the discharge transitions into the pull phase, in which the current sheet changes both its orientation and polarity. The pull current sheet has negative $J_{\theta}$ and is elongated mainly along the $Z$ direction. In contrast to the push phase, the $B_{\theta}$ profile during the pull phase exhibits a clear quadrupole structure \citep{ren05, ren08}, which is a well-known signature of two-fluid effects in magnetic reconnection. In these discharges, the OH coils are also used to supply additional toroidal current, and we observe an overall increase in electron temperature of approximately $\sim 50$ \%, indicating the effectiveness of the OH system in strengthening the reconnection drive.

\begin{figure}
    \centering
    \includegraphics[width=1\linewidth]{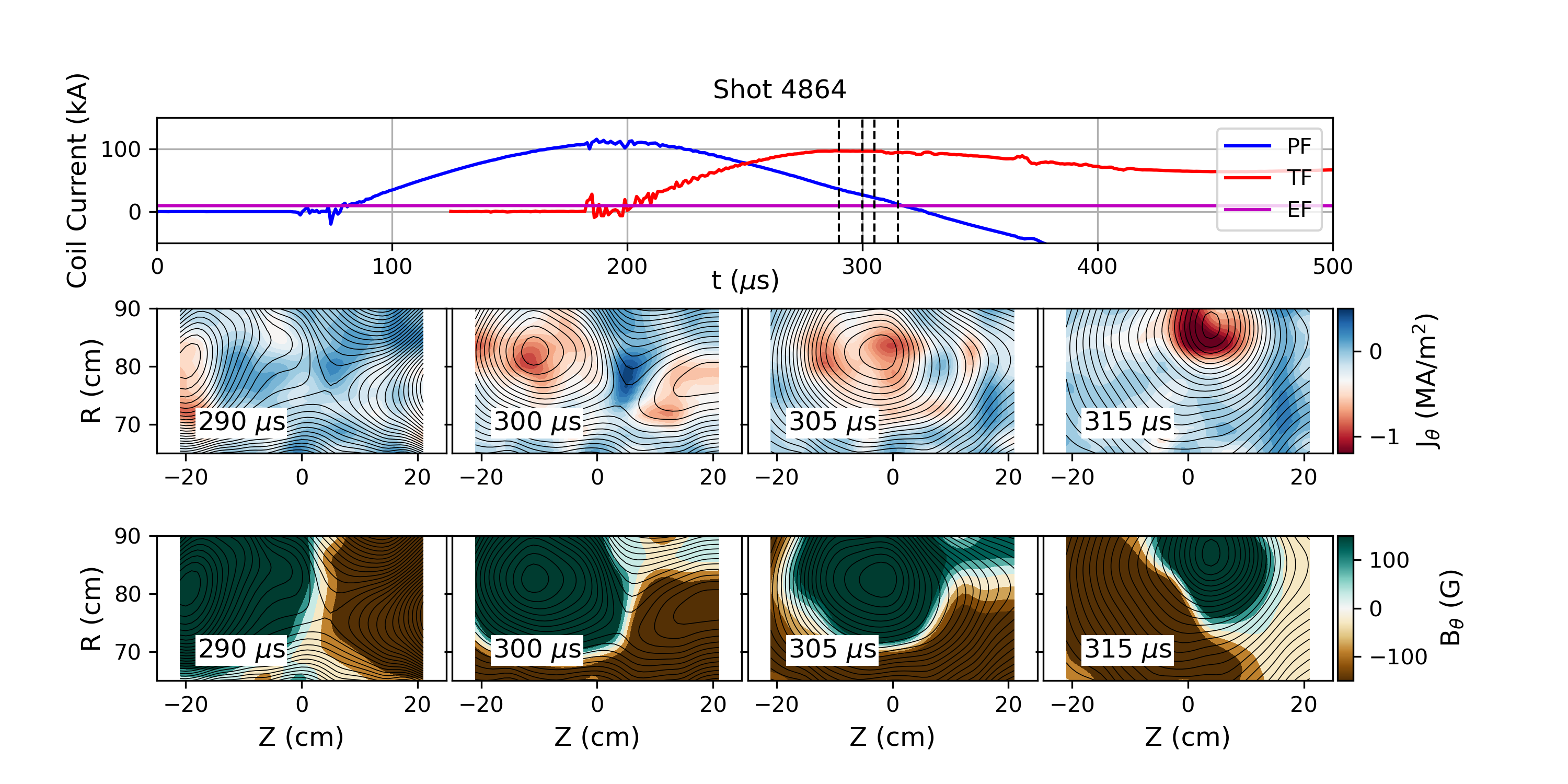}
    \caption{Example of a merging discharge in FLARE using argon in counter-helicity, Case-I operation. The top panel shows the PF, TF, and EF coil currents; the delayed crowbar of the PF bank causes the PF current to reverse sign, leading to pinch-off of the plasmas near the two flux cores. The middle panels show the evolution of the out-of-plane current density $J_{\theta}$, and the bottom panels show the corresponding out-of-plane magnetic field $B_{\theta}$, demonstrating the merging process of two spheromak-like plasmas to form a reversed-field-pinch-like configuration and eventually relax to a field-reversed configuration. Black lines are poloidal flux contours.}
    \label{fig:merging}
\end{figure}

The second example, shown in Fig.~\ref{fig:merging}, demonstrates a merging configuration. In this case, the PF capacitor banks are crowbarred at a later time so that the PF coil current reverses sign, as shown in the top panel. Argon is used as the working gas in this discharge to improve plasma stability \citep{gerhardt07}. The negative PF current causes the plasma near each flux core to pinch off, leading to the formation of two spheromak-like plasmas that subsequently move toward each other and merge. 

The middle panels show the evolution of the out-of-plane current density, $-J_{\theta}$, and the bottom panels show the corresponding out-of-plane magnetic field, $B_{\theta}$. At $t = 290~\mu$s, two pinch-off spheromaks are clearly visible and begin to interact with each other. By $t = 300~\mu$s, a push-type current sheet forms between the two spheromaks, indicating active reconnection during the merging process. At $t = 305~\mu$s, the two plasmas have largely merged into a single configuration. Unlike the corresponding MRX merging case \citep{gerhardt06,gerhardt07}, the toroidal magnetic-field component remains substantial after merging and continues to show a clear reversal from the center toward the edge. As a result, the merged plasma takes the form of a reversed-field pinch (RFP) \citep{prager05}, rather than a field-reversed configuration (FRC) \citep{gerhardt06}. At $t = 315~\mu$s, further self-organization is observed, during which the magnetic axis shifts radially outward. 

This behavior suggests a distinct relaxation pathway in FLARE compared with earlier merging experiments and highlights the rich dynamics enabled by its larger scale and flexible operating space. The total poloidal flux in the merged plasma is estimated to exceed 25~mWb. A more detailed investigation of the merging dynamics, magnetic relaxation, and associated reconnection physics will be an important subject for future work.

\begin{figure}
    \centering
    \includegraphics[width=1\linewidth]{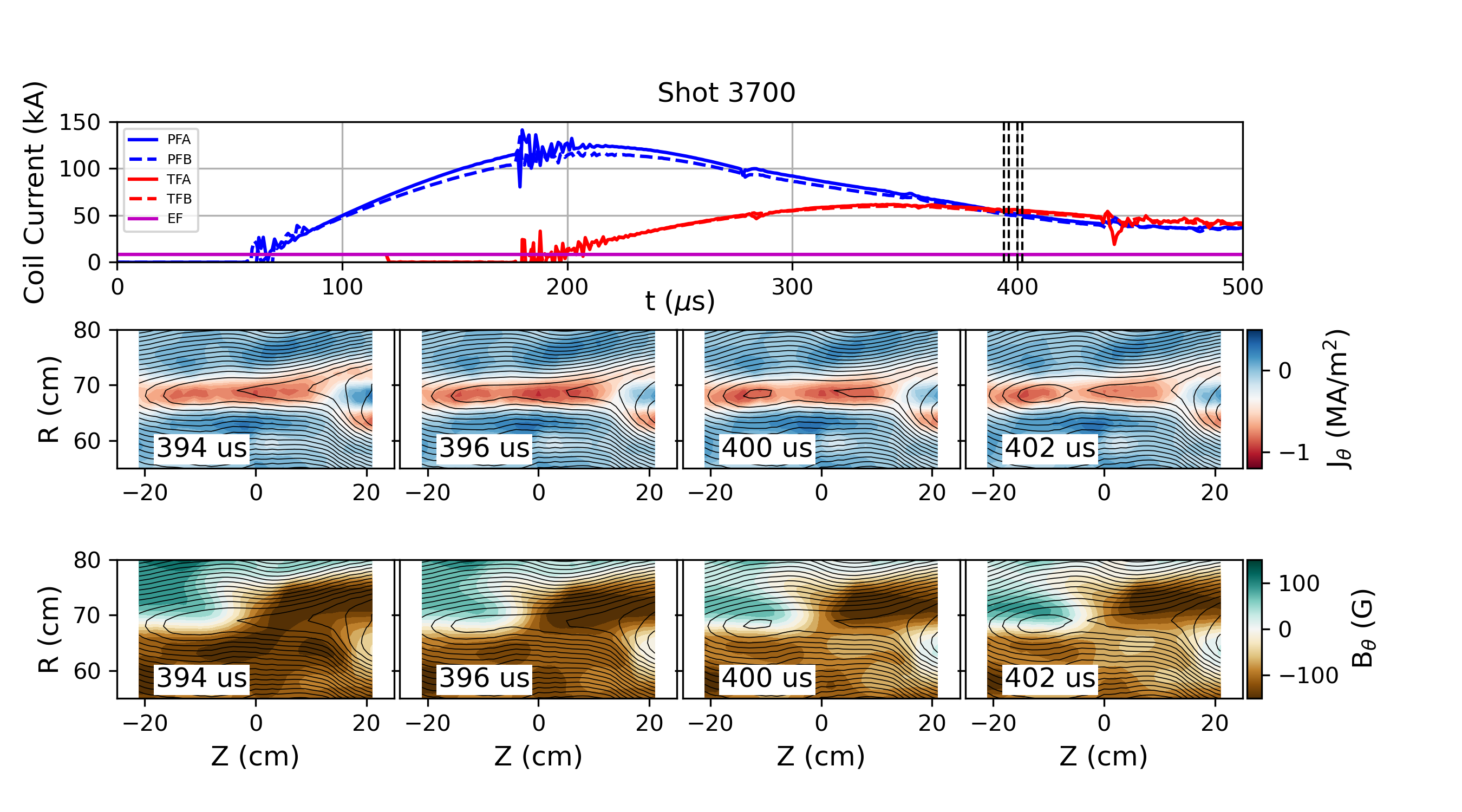}
    \caption{Example of an asymmetric downstream reconnection discharge in FLARE using helium, produced by independently controlling the two PF capacitor banks. The top panel shows the PF, TF, and EF coil currents; the PFA current (right side) is intentionally larger than the PFB current (left side), producing a left--right asymmetry in the downstream magnetic pressure. The middle panels show the evolution of the out-of-plane current density $J_{\theta}$, and the bottom panels show the corresponding out-of-plane magnetic field $B_{\theta}$, at four representative times. Black lines are poloidal flux contours.}
    \label{fig:asym}
\end{figure}

The third example, shown in Fig.~\ref{fig:asym}, demonstrates an asymmetric downstream configuration using helium as the working gas and produced by independently controlling the two PF capacitor banks. In this discharge, the current waveforms of PFA (right side) and PFB (left side) are intentionally made different by charging the PFA bank to a voltage approximately 10 \% higher than that of PFB. Although this difference in the drive current is relatively small, it produces a pronounced left-right asymmetry in the resulting reconnection geometry. 

The physical origin of this downstream asymmetry is that the side with the higher PF current, namely the right side, stores more magnetic energy and therefore exerts a stronger magnetic pressure, analogous to the earthward side of the terrestrial magnetotail. This imbalance forces the reconnection outflow to develop asymmetrically. In particular, the structures observed around $t = 400~\mu$s and $402~\mu$s resemble the geometry of an asymmetric solar-flare eruption, in which one downstream side is more strongly confined while the other side opens more readily. 

The middle panels show that the current-sheet structure becomes strongly distorted, while the bottom panels reveal a corresponding asymmetry in the out-of-plane magnetic field, $B_{\theta}$. In particular, the right side does not exhibit a clear negative $B_{\theta}$ region, whereas the left side shows both positive and negative $B_{\theta}$, forming a more evident quadrupole-like structure. This suggests that the left side, where the downstream magnetic pressure is weaker, is more favorable for the development of reconnection outflow. To further test this interpretation, future experiments will directly measure the ion outflow profiles using Mach probes under these asymmetric operating conditions and determine whether the asymmetric $B_{\theta}$ structure is accompanied by a corresponding asymmetry in the ion exhaust.

Together, these examples highlight the versatility of FLARE in generating a wide range of reconnection scenarios with precise control over boundary conditions and drive parameters. The combined use of PF, TF, EF, GF, and OH systems, together with the capability for independent control of the two flux cores, enables systematic exploration of reconnection physics across multiple regimes, including symmetric push-pull operation, spheromak merging, and asymmetric downstream outflow.

\section{Future Upgrades and Collaborative Research}
The successful commissioning and initial operation of the Facility for Laboratory Reconnection Experiments (FLARE) mark the beginning of a broad research program. To fully exploit the machine’s capabilities and access more extreme reconnection regimes, a series of engineering, diagnostic, and computational upgrades are planned for the near future, together with a transition toward collaborative user operations.

To expand the accessible parameter space toward higher Lundquist numbers and larger normalized system sizes, several important engineering upgrades are planned following the inaugural experimental campaign to achieve the Stage 3 capabilities shown in Fig.~\ref{fig:stages}. One major upgrade is the installation of the additional drive coils described above, which will enable stronger and more versatile inductive flux injection. Another priority is enhanced structural reinforcement of the guide-field (GF) coil system. This upgrade will allow FLARE to safely sustain higher guide fields of up to $0.5$ T, enabling detailed studies of strong-guide-field reconnection relevant to magnetically confined fusion plasmas, as well as multiple-X-line dynamics and particle acceleration. 

Although the initial diagnostic suite, including magnetic-probe arrays, Langmuir probes, and interferometry, already provides essential measurements of the global and local plasma properties, a more complete understanding of kinetic reconnection requires substantially expanded diagnostic capability. Over the next several years, FLARE will integrate a number of advanced diagnostics to provide a more comprehensive picture of the reconnection layer.

\begin{figure}
    \centering
    \includegraphics[width=1\linewidth]{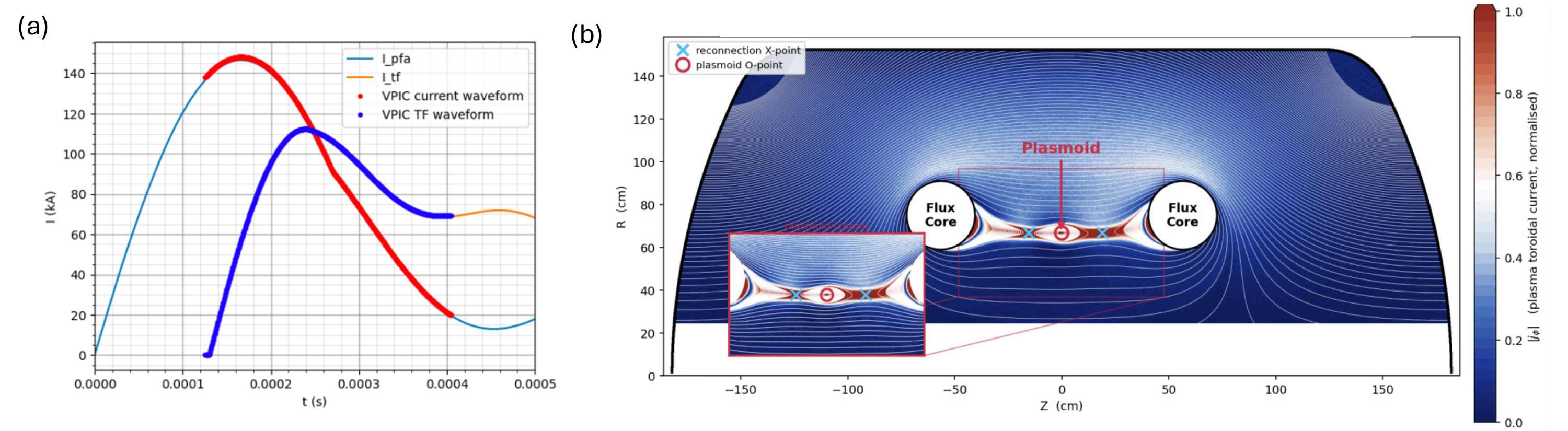}
    \caption{Example of a fully kinetic Vector Particle-in-Cell (VPIC) simulation designed to model FLARE discharges. (a) Temporal evolution of the PF and TF coil currents. To accurately bridge laboratory experiments with numerical modeling, the simulation utilizes realistic coil currents, with the simulated PF (VPIC PF current waveform, red dots) and TF (VPIC TF waveform, blue dots) precisely tracking the measured current waveforms (solid lines). (b) A two-dimensional snapshot of the normalized plasma toroidal current overlaid with poloidal magnetic flux surfaces (white contours) during the reconnection phase. The simulation incorporates the accurate macroscopic cylindrical geometry and physical flux core locations. A zoomed-in inset highlights the fine-scale structure of the reconnecting current sheet, explicitly capturing the formation of a central plasmoid (O-point, red circle) bounded by multiple reconnection X-points (blue crosses), which is characteristic of the multiple X-line regime. }
    \label{fig:vpic}
\end{figure}

Because numerical models often struggle to capture the multiscale nature of magnetic reconnection in large space and astrophysical plasmas, the integration of experimental measurements with advanced simulations is a central element of the FLARE program. State-of-the-art fully kinetic particle-in-cell (VPIC) simulations under cylindrical geometry will be carried out in parallel with the experimental campaigns to test theoretical models and interpret the observed dynamics \citep{bowers08}. These simulations include a first-principles model of Coulomb collisions and they can be used to model FLARE current sheet dynamics from the collisionless to collisional regimes. 

As an initial demonstration of this capability, Fig. \ref{fig:vpic} presents a baseline VPIC simulation modeled directly on the macroscopic cylindrical geometry of the FLARE vacuum vessel. To accurately replicate the experimental conditions, the numerical model employs realistic PF and TF coil currents. As shown in the resulting two-dimensional cross-section, this fully kinetic treatment successfully captures the global magnetic topology. Crucially, the high-resolution inset reveals the fine-scale dynamics of the thinning current sheet, explicitly identifying a central plasmoid (magnetic O-point) flanked by multiple reconnection X-points. This confirms the simulation's ability to resolve the highly dynamic multiple X-line regime, making these tightly coupled numerical studies vital for bridging the gap between laboratory-scale reconnection in FLARE and the much larger scales relevant to heliophysical and astrophysical systems.

Following its initial operational phase, FLARE is expected to evolve into a U.S. Department of Energy collaborative research facility, open to researchers from universities, national laboratories, and private industry. The scientific direction of the facility will be guided by a Science Advisory Board, while access to experimental runtime will be managed through a Facility Scheduling Committee based on the scientific merit and technical feasibility of user proposals.

To support external users, PPPL plans to organize dedicated user-support and research teams spanning space physics, solar and astrophysical plasmas, and discovery plasma science and fusion. FLARE is also expected to connect with broader community efforts, including collaborative networks such as MagNetUS \citep{hurst25}, in order to streamline scientific coordination and broaden user engagement. Through this open and collaborative model, FLARE will serve not only as a unique platform for discovery plasma science, but also as a foundation for broader multi-institutional and public-private partnerships aimed at advancing the understanding of magnetic reconnection across multiple disciplines.

Finally, we conclude this facility paper by providing an initial list of research topics for FLARE based on the major scientific challenges in magnetic reconnection, as summarized in recent community whitepapers~\citep{ji20,ji23c} submitted to several decadal surveys. Each of the ten major problems contains a set of sub-problems which are listed below as questions for further reference:

\begin{enumerate}
\item \textit{Multiple scale problem}: How does reconnection couple global fluid (magnetohydrodynamic or MHD) scales to local dissipation (kinetic) scales? Does plasmoid instability exist in MHD or kinetic plasmas? If so, what are the critical Lundquist number and normalized system size? How do the resultant multiple X-lines and the corresponding reconnection rate scale with controlling parameters in each regime? What are effects due to asymmetry in upstream and/or downstream and a finite guide field?
\item \textit{3D problem}: How does reconnection take place in 3D on global and local scales? What are possible forms of plasmoid instability in 3D, and do they evolve into interacting flux ropes? How does multiple X-line reconnection scale with the third dimension, and can it become turbulent? Is it possible to externally drive tearing mode reconnection? How do multiple tearing modes interact, and can they lead to magnetic stochasticity? 
\item \textit{Energy problem}: How are particles heated and accelerated? How do particle heating and acceleration scale with the Lundquist number and normalized system size? What are the apportionments between electrons and ions which can have multiple components? What are effects due to a finite guide field?
\item \textit{Boundary problem}: How do boundary conditions affect the reconnection process? How does downstream magnetic and/or plasma pressure affect multiple X-line reconnection? What are the effects due to line-tied boundaries in the third direction?
\item \textit{Onset problem}: How does reconnection start? Is reconnection onset local or global? Does reconnection start simultaneously in 2D or essentially in 3D, and how does it depend on multiple X-lines?
\item \textit{Partial ionization problem}: How does partial ionization affect reconnection? Do neutrals promote or suppress multiple-scale reconnection? How, and to what extent, are neutral particles energized during reconnection?
\item \textit{Explosive phenomena problem}: How, and under what conditions, does magnetic reconnection act either as a driver or a consequence of explosive phenomena such as solar flares, Coronal Mass Ejections, and disruptive phenomena in fusion plasmas? Are solar eruptions triggered by torus instability, kink instability, or a combination of both? How does reconnection contribute to these eruptions?
\item \textit{Flow-driven problem}: What role does reconnection play in dynamos of flow-driven plasmas? Does reconnection saturate the dynamo by dissipating magnetic energy or facilitate dynamo by relaxing its topological constraints?
\item \textit{Turbulence and shock problem}: What role does reconnection play in related processes such as turbulence, collisionless shocks, and plasma transport? Is reconnection an integral part of magnetized turbulence? Is turbulence a solution for multi-scale reconnection? How does reconnection facilitate particle acceleration in collisionless shocks? 
\item \textit{Extreme condition problem}: How does reconnection operate in extreme conditions such as intense radiation, relativistic environments, and even quantum electrodynamic (QED) effects, as seen in astrophysics?
\end{enumerate}

Building on the successes of the past two decades~\citep{yamada10,ji23}, the FLARE facility will provide the community with rich research opportunities to solve a majority of these problems, especially the first seven, under well-controlled and well-diagnosed conditions in a laboratory setting. Tackling the last three problems, however, requires additional means and platforms. Flow-driven reconnection can be realized in high-energy-density (HED) plasmas generated either by high-power lasers~\citep[e.g.][]{nilson06,li07,willingale10,zhong10,fiksel14,chien23} or by pulsed power~\citep{hare17}, and sometimes with turbulence~\citep{ping23}, intense radiation~\citep{datta24}, and even electron-positron plasmas~\citep{russell26}. Co-locating FLARE with a high-power laser and/or a pulsed power system would significantly broaden the accessible parameter space to address the flow-driven problem as well as turbulence and shock problem.

\section*{Acknowledgments}
The authors wish to thank a long list of colleagues who contributed to and supported this project directly or indirectly from various directions since 2013, including Vassilis Angelopoulos, JP Allain, Spiro Antiochos, Ron Bell, Elena Belova, Scott Bentivegna, Mike Brown, Joerg Buechner, Jim Burch, Andy Carpe, Fausto Cattaneo, Li-Jen Chen, Adam Cohen, Steven Cowley, Daren Craig, Susan Dawson, Pablo Debenedetti, Bart De Pontieu, Ed DeLuca, Ahmed Diallo, Peng E, Fatima Ebrahimi, Eric Edlund, Philip Efthimion, Chris Eisgruber, Robert Ergun, Gennady Fiksel, Kristen Fischer, Mike Ford, Cary Forest, Darren Garnier, Charlie Gentile, Kris Gilton, Steve Gitomer, Aaron Goodman, Jeremy Goodman, Olaf Grulke, Fan Guo, Ke Han, Jack Hare, Rich Hawryluk, Michael Hesse, Les Hill, Scott Hsu, Yi-Ming Huang, Tom Intrator, Marie Iseicz, Frank Jenko, Austin Jones, Bill Jones, Therese Jorgensen, Homa Karimabadi, Judy Karpen, Justin Kasper, Jim Kukon, Russell Kulsrud, Matt Kunz, Ari Le, Ben LeBlanc, Tony Lee, Hui Li, Jiangang Li, Yu Lin, Yi-Hsin Liu, Nuno Loureiro, Quanming Lu, Tim Luce, Slava Lukin, Steve Majeski, Dave McComas, Paul Melnik, Tim Meyer, Forrest Mozer, Nick Murphy, David Pace, Lyman Page, Luke Perkins, Tai Phan, Nirmol Podder, Mario Podesta, Nikolai Pogorelov, Deborah Prentice, Jackie Pursell, Roger Raman, Jennifer Redford, Valeria Riccardo, Vadim Roytershteyn, Robert Rosner, Guy Rossi, Peter Schiffer, Earl Scime, Lorenzo Sironi, Mikhail Sitnov, Peter Sloboda, Stew Smith, Yuntao Song, David Spergel, Anatoly Spitkovsky, Tim Stevenson, Michael Strauss, Jim Stone, Francine Taylor, Tim Tharp, Ed Thomas, Stefanie Thomas, Peter Titus, Roy Torbert, Dmitri Uzdensky, Setthivoine You, Jim Van Dam, Marco Velli, Herman Verlinde, Mike Viola, Haimin Wang, Xiaogang Wang, Simon Woodruff, Ali Yazdani, Brenda Zanghi, Gary Zank, Mike Zarnstorff, Jinxing Zheng, and Ellen Zweibel. In addition, more than 100 authors of multiple community whitepapers~\citep[e.g.][]{ji23c} are acknowledged for their support and interest. Finally, we acknowledge a group of plasma graduate students including Dennis Boyle, Lee Ellison, Craig Jacobson, Clayton Myers, Jeff Parker, Jacob Schwartz, and Jono Squire, who collectively conceived the FLARE name in 2014.

\section*{Funding}
The FLARE facility was developed through funding from the U.S. Department of Energy under contract \# DE-AC02-09CH11466 and Princeton University. The initial FLARE instrument construction was funded by the U.S. National Science Foundation Major Research Instrument grant \# PHY-1337831 and Princeton University, the University of Wisconsin-Madison, and the University of Maryland. The FLARE VPIC simulation model was developed through funding from the U.S. Department of Energy Fusion Energy Sciences Research in Basic Plasma Science and Engineering FWP LANLE8A6.

\bibliographystyle{jpp}
\bibliography{reconnection,misc}

\end{document}